\RequirePackage{fix-cm}
\documentclass[natbib,twocolumn,final]{svjour3}          % twocolumn
\usepackage{soul}
\usepackage{amssymb}
\usepackage{hyperref}
\usepackage{mathtools, cuted}
\usepackage{commath}
\usepackage{lineno}
\usepackage{epsf,epstopdf}
\usepackage{siunitx}
\usepackage{graphicx,amssymb,amsmath}
\usepackage{txfonts}
\usepackage{natbib}
\usepackage{mathptmx}      % use Times fonts if available on your TeX system
\usepackage{aps-bibstyle}  % use this style if you upload to .tex file only a part of Bibtex created bbl.
\usepackage[first=0,last=9]{lcg}
\usepackage[table]{xcolor}
\usepackage{url}
\hypersetup{pdftitle = The title of my PDF, pdfauthor = My name, pdfsubject= The subject, pdfkeywords = keyword1 keyword2 keyword3}
\hypersetup{colorlinks = true, linkcolor = blue, anchorcolor = red, citecolor = blue, filecolor = red, pagecolor = red, urlcolor = blue}
 \journalname{to be inserted}
\newcommand{\aap}{{Astron. Astrophys.}}

\begin{document}
%%%%%%%%%%%%%%%%%%%%%%%%%%%%%%%%%%%%%%%%%%%%%%%%%%%%%%%%%%%%%%%%%%%%%%%%
\title{On the Spatial Collinear Restricted Four-Body Problem With Non-Spherical  Primaries}
%\subtitle{The analysis of periodic orbit in the restricted four-body problem using mobile coordinates}
\author{Md Sanam Suraj  \and
        Rajiv Aggarwal       \and
              Amit Mittal\and
              Om Prakash Meena\and
              Md Chand Asique
}
\institute{Md Sanam Suraj \at
    Department of Mathematics,
    Sri Aurobindo College, University of Delhi,  New Delhi-110017, Delhi, India\\
    \email{\url{mdsanamsuraj@gmail.com}}\\
    \email{\url{mdsanamsuraj@aurobindo.du.ac.in}}           %  \\
           \and
  Rajiv Aggarwal \at
  Department of Mathematics,
  Deshbandhu College, University of Delhi, New Delhi-110019, Delhi, India\\
              \email{\url{rajiv_agg1973@yahoo.com}}
%%%%%%%%%%%%%%%%%%%%%%%%%%%%%%%%%%%%%%%%%%%%%%%%%%%%%
\and
Amit Mittal\at
Department of Mathematics,
ARSD College, University of Delhi, New Delhi-110021, Delhi,  India\\
 \email{\url{to.amitmittal@gmail.com}}
 %%%%%%%%%%%%%%%%%%%%%%%%%%%%%%%%%%%%%%%%%%%%%%%%%%%%%
 \and
 Om Prakash Meena \at
  Department of Mathematics,
  Deshbandhu College, University of Delhi, New Delhi-110019, Delhi, India\\
\and
Md Chand Asique\at
Deshbandhu College, University of Delhi, New Delhi-110019, Delhi,  India\\
 \email{\url{mdchandasique@gmail.com}}
 }
\date{Received:     date / Accepted:             date}
%\date{today}
\maketitle
\begin{abstract}
In the present work a systematic study has been presented in the context of the existence of libration points, their linear stability, the regions of motion where the third particle can orbit and the domain of basins of convergence linked to libration points in the spatial configuration of the collinear restricted four-body problem with non-spherical primaries (i.e., the primaries are oblate or prolate spheroid). The parametric evolution of the positions of the libration points as function of the oblateness and  prolateness parameters of the primaries and the stability of these points in linear sense are illustrated numerically.  Moreover, the numerical investigation shows that the only libration points  which lie on either of the axes are linearly stable for several combinations of the oblateness parameter and mass parameter whereas the non-collinear libration points are found linearly unstable, consequently unstable in nonlinear sense also,  for studied value of mass parameter and oblateness parameter. Moreover, the regions of possible motion are also depicted, where the infinitesimal mass is free to orbit, as function of Jacobian constant. In addition, the basins of convergence (BoC) linked to the libration points are illustrated by using the multivariate version of the Newton-Raphson (NR) iterative scheme.
\end{abstract}
%%%%%%%%%%%%%%%%%%%%%%%%%%%%%%%%%%%%%%%%%%%%%%%%%%%%%%%%%%%%%%%%%%%%%%%%
%%%%%%%%%%%%%%%%%%%%%%%%%%%%%%%%%%%%%%%%%%%%%%%%%%%%%%%%%%%%%%%%%%%%%%%%%%%
%\phantomsection
%\addcontentsline{toc}{section}{Introduction}
\section{Introduction}
The restricted problem of the four bodies with various perturbations have been fascinated by a large group of astronomers, researchers and scientists over the globe in few decades. The restricted problem of four bodies in planar case is referred as restricted $(N+1)$ where $N=3$, body problem. In this problem the fourth body does not influence  the motion of the three primaries, and consequently the fourth particle can be assumed as a dynamical system composed of an infinitesimal mass (i.e., the test particle) together with the three main primaries.  The central configuration of the three-body problem contains two type of configurations:  the Euler configuration (i.e., straight-line configuration) and the Lagrange configuration (i.e., the equilateral triangle configuration) (see \cite{L06}, \cite{H16}). The classical restricted three-body problem with various perturbations has been studied by many researchers in the few decades. For example the existence as well as stability of the equilibrium points (e.g., \cite{SMB85}, \cite{A12}, \cite{AA19a}, \cite{SGA19}), when the primaries are non spherical in shape (e.g., \cite{SS79}, \cite{AGS16a}), the analysis of periodic orbits (e.g., \cite{AGL19b} ), the basins of convergence and classifications of orbits (e.g., \cite{Z15a}, \cite{Z15b}) have been studied in restricted three-body problem.

In the few decades, the Lagrange and Euler configurations of the planar restricted four-body problem (PR4BP)  have been discussed by various researchers by modifying the effective potential by adding some extra terms which occur due to various perturbations to unveil the existence of equilibrium points and their stability (e.g.,  \cite{M81}, \cite{KAE06}, \cite{MP08}, \cite{BP11},  \cite{BCV17},  \cite{SAA18}), the study of the motion of the test particle when its mass is variable (e.g., \cite{SAA19}, \cite{SMC18}), the computations of periodic orbits (e.g., \cite{BP11b}, \cite{PAA19} ), the study of fractal basins of convergence (e.g.,  \cite{Z17}, \cite{SZK18}), or the study of orbital dynamics of escape and collisions (e.g., \cite{Z16}, \cite{M98}).

In various research papers, the scientists have also studied the  existence of libration points and their stability  in the same dynamical model by taking the shape of primaries. \cite{APH15} have studied the existence of libration points in the  photogravitational version of the restricted four-body problem when one of the primary is an oblate/prolate spheroid, whereas   \cite{APH16} have discussed the same model by taking a triaxial rigid body as primary. \cite{KK12} have included the effect of  solar wind drag to discuss the libration points  and zero velocity surfaces in the restricted problem of four bodies. In Ref. \cite{Zo15c}, the author has compared the orbital dynamics in three models which describe the various properties of a star cluster  which  rotates in a circular orbit around its parent galaxy. Moreover, \cite{ZJ18} have discussed the orbit and escape dynamics in barred galaxies and illustrated the basins of escape linked  to the escape through the escape channels in the vicinity of  the libration points $L_{2,3}$ which are symmetrical in nature  and further established the relation with the corresponding distribution of the escape times of the orbits. In \cite{ZDG18}, the motion of the test particle in a non-spinning binary black hole system is investigated numerically when the masses are equal. Further they have classified the initial conditions into three different categories, i.e., bounded, escaping and displaying close encounters by using the smaller alignment index  chaos indicator, the orbits are classified into regular, sticky and chaotic.

In the literature, there is plethora of papers  available where the spatial collinear restricted four-body problem (CR4BP) has been discussed. Ref. \cite{AAE16a}, have discussed the planar motion of the infinitesimal body moving under the system of three main primaries situated in a collinear configuration (Euler 1767) with a symmetry and the peripheral primaries are also source of radiation. In addition, they have considered the case where the gravitational force is less than the radiation force. In the mentioned setup, they have discussed the analytic study of the position and stability of the libration points for the involve parameters. In continuation of their study,  Ref. \cite{AAE16b} have discussed the out-of-plane libration points, i.e., the equilibrium points which lie out side the configuration plane of primaries in the same dynamical system. \cite{BCV17} have discussed the  relative equilibria and their stability of the infinitesimal mass,  moving under the gravitational attraction of three primaries which are  in a syzygy,  under the  repulsive Manev potential.  \cite{Z17b} has investigated the BoC linked with the equilibrium points by applying the NR iterative scheme in the collinear restricted four-body problem with angular velocity.  He has illustrated the parametric evolution of the positions of equilibrium points and their stability  when the values of the mass parameter and the angular velocity are considered in the fixed intervals. \cite{PAA19} have illustrated the  evolution of families of symmetric periodic orbits as the function of mass parameter and evolution of spiral points, which show the connection between heteroclinic orbits and equilibrium points. \cite{ABM19} have illustrated the  high order parametrizations of the stable and unstable manifolds connected  with the  libration  points to allocate the ejection orbits that eject from quadruple collision, in addition,  they have also shown the existence of ejection-direct escape orbits analytically.

In the present manuscript we proposed to discuss the motion of the infinitesimal mass, moving  in the gravitational influence of the main primaries which are in a syzygy, this setup always referred as collinear restricted four-body problem. In this study we extend the work of authors \cite{PAA19}  by taking the peripheral primaries  as oblate/prolate spheroid.

The present manuscript has following structure: the description of the mathematical model and the equations of motion of the test particle are discussed in Sec. \ref{Sec:2}. The Sec. \ref{LP:0}  deal with the existence of the libration points as function of parameters whereas stability of these libration points is analyzed in the following section. The regions of possible motions are presented in Sec. \ref{ZVC:0}. The BoC linked with the libration points are illustrated in Sec \ref{BoC}. The  paper ends with Sec. \ref{CR:0} where the concluding remarks are presented.
%%%%%%%%%%%%%%%The  Description of Mathematical Model%%%%%%%%%%%%%%%%%%%
\section{Description of Mathematical model and Equations of motion}
\label{Sec:2}
The dynamics of the infinitesimal mass (also referred as test particle) moving under the gravitational influence of the three primaries $P_i, i=0,1,2$ has been investigated in the present problem. It is assumed that the primaries $P_1$ and $P_2$,  having the same mass $m_1=m_2=m$, are oblate/prolate bodies and the oblateness/prolateness  parameters $A_1=A_2=A$ are also same while the central primary $P_0$ is spherical in shape. Moreover, we have assumed that the primaries are in collinear central configuration where the primaries $P_1$ and $P_2$ are situated symmetrically with respect to the central primary $P_0$. Also $\beta m$, taken as the mass of the central primary $P_0$ where $\beta$ denotes the so called mass parameter,  is placed at the center of masses of the system which is taken as origin of the reference frame. The peripheral bodies $P_{1,2}$ describe the circular orbit around the central primary $P_0$ with same angular velocity $\omega$.

In the synodic frame of reference with the same origin, the line passes through origin and joining the center of mass of $P_1$ and $P_2$ is assumed as $x-$axis and $Oy-$axis  is a line perpendicular to $Ox-$axis and passing through origin while the line through origin and perpendicular to instantaneous plane of the primaries is assumed as $Oz-$axis. Consequently, in the synodic frame of reference, the coordinates of the primaries $P_i,  i=0,1,2$ are $(x_i, 0, 0)$,  $x_0=0, x_1=-x_2=\frac{1}{2}$ , and $Gm_0=\beta$.
The configuration of the primaries remains invariant, if the sum of the total  gravitational force exerted by $P_0$ and $P_2$ on $P_1$  are equal to the centrifugal force, i.e.,
\begin{eqnarray}\label{Eq:00}
 m_1\omega^2\|P_0P_1\|& =&\frac{Gm_1m_0}{\|P_0P_1\|^2} + \frac{3Gm_0}{2\|P_0P_1\|^4}I^* + \frac{Gm_1m_2}{\|P_1P_2\|^2} \nonumber\\
 &&+\frac{3Gm_2}{2\|P_1P_2\|^4}I^*+\frac{3Gm_1}{2\|P_1P_2\|^4}I^{'*},\nonumber
\end{eqnarray}
where
\begin{eqnarray}
  I^* &=& (I_1+I_2+I_3-3I),\quad
  I^{'*}=(I_1^{'}+I_2^{'}+I_3^{'}-3I^{'}), \nonumber
\end{eqnarray}
whereas $I_i$ and $I_i^{'}, i=1,2,3$  are the principal moment of inertia of the primaries $P_1$ and $P_2$ at its center of mass $O$, respectively while $I$ is the moment of inertia  about the line joining the center of mass of the primary $P_1$ and the primary $P_0$ or $P_2$ and $I^{'}$ is the moment of inertia about the line joining the center of mass of the primary $P_2$ and the primary $P_0$ or $P_1$. Therefore, the angular velocity of the synodic frame is
\begin{equation}\label{E:001}
 \omega^2=\Lambda=2(1+4\beta)+6A(1+8\beta),
\end{equation}
where  $ A_1 =\frac{a_1^2 - c_1^2}{5R^2},   A_2=\frac{a_2^2 - c_2^2}{5R^2}$ and in the Copenhagen case we have taken $A_1=A_2=A$.
Also, when the oblateness of the primaries are neglected i.e., $A=0$, the $\omega^2$ reduces to same as in Eq.(1) in Ref. \cite{AAE16b}.

%%%%%%%%%%%%%%%%%%%%%%%%%%%%%%%%%%%%%%%%%%%%
\begin{figure}%
\centering
\resizebox{\hsize}{!}{\includegraphics{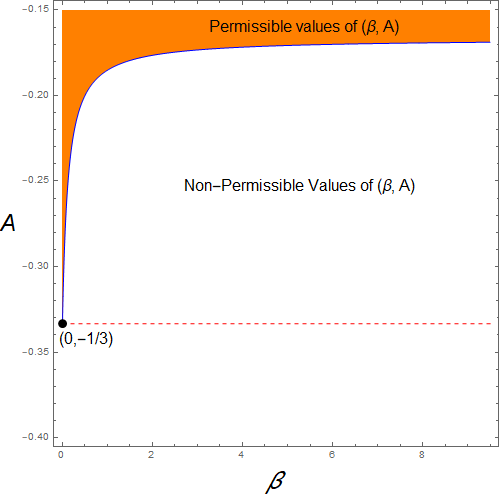}}
\caption{The bifurcation curve in the $(\beta, A)$ plane.  (colour figure online).}
\label{Fig:01N}
\end{figure}
%%%%%%%%%%%%%%%%%%%%%%%%%%%%%%%%%%%%%%%%%%%%%%%%%%
The units of distances, mass and time are chosen in such a way that $\|P_1P_2\|=1$, and $Gm=1$.
The equations of motion of the test particle after a change of time $ds=\omega dt$, where both of the peripheral bodies are oblate, can be written as:
\begin{subequations}
\begin{eqnarray}
\label{Eq:1a}
\ddot{x}-2\dot{y}&=&\frac{\partial U}{\partial x},\\
\label{Eq:1b}
\ddot{y}+2\dot{x}&=&\frac{\partial U}{\partial y},\\
\label{Eq:1c}
\ddot{z}&=&\frac{\partial U}{\partial z},
\end{eqnarray}
\end{subequations}
where
\begin{eqnarray*}
  U(x, y,z)&=& \frac{1}{2}(x^2 + y^2) + \frac{1}{\Lambda}\bigg\{\frac{\beta}{r_0} + \sum_{i=1}^{2}\Big(\frac{1}{r_i}+\frac{A_i}{2 r_i^3}\Big) \bigg\},\nonumber\\
  r_i&=&\sqrt{(x-x_i)^2+y^2+z^2}, i=0,1,2.
\end{eqnarray*}

Furthermore, analogous to the  restricted problem of three bodies, the system of equations in Eqs. \ref{Eq:1a}-\ref{Eq:1c} possesses the first integral, always considered as Jacobi integral, and expressed as
\begin{equation}\label{Eq:3N}
  C=2U(x, y, z)-(\dot{x}^2+\dot{y}^2+\dot{z}^2),
\end{equation}
where the Jacobian constant is represented by $C$.
%%%
%%%%%%%%%%%%%%%%%%%%%%%%%%%%%%%%%%%%%%%%%%%%
\begin{figure*}%
\centering
\resizebox{\hsize}{!}{\includegraphics{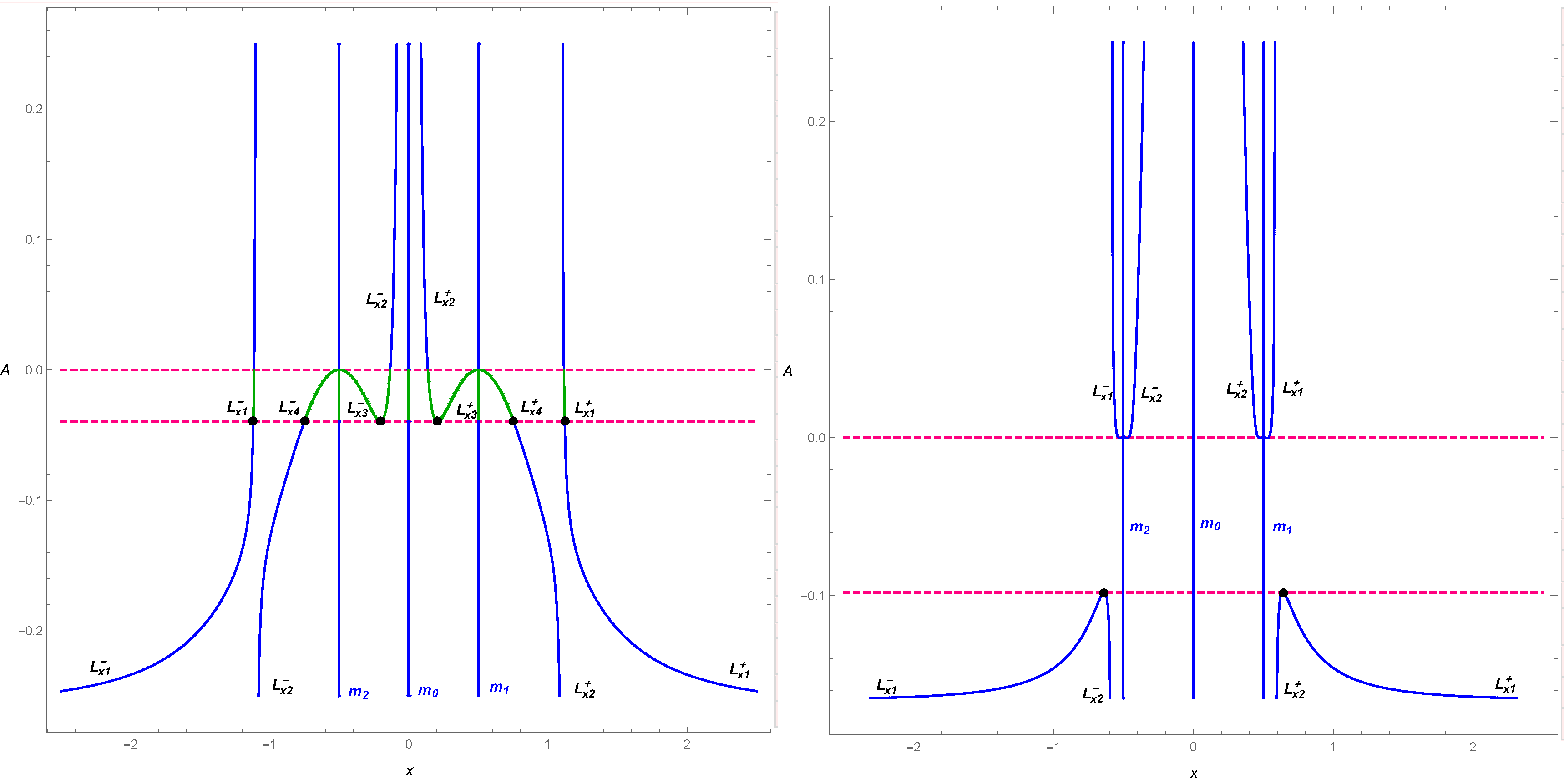}}
\caption{ Evolution of the positions of collinear libration points $L^+_{x1,2,3,4}$ and $L^-_{x1,2,3,4}$ as function of parameter $A$; (a) \emph{left}: for the fixed value of $\beta=0.1$. The horizontal dashed \emph{magenta} lines show the critical value of $A=0$ and $A=-0.039515$, for this interval of $A$, there exist eight collinear libration points.  (b) \emph{right}: for $\beta=1000$ and the horizontal dashed \emph{magenta} lines show the critical value of $A=-0.09805989$, for this interval of $A \in (-0.166687, -0.09805989)$ and $(0, 0.25)$ , there exist four collinear libration points whereas in $A\in(-0.09805989,0)$ there exist no collinear libration points.  (colour figure online).}
\label{Fig:1N}
\end{figure*}
%%%%%%%%%%%%%%%%%%%%%%%%%%%%%%%%%%%%%%%%%%%%%%%%%%
\section{The libration points}
\label{LP:0}
The libration points can be obtained by solving the equations
\begin{equation}\label{Eq:}
  U_x=0, U_y=0, U_z=0,
\end{equation}
where $U_x, U_y, U_z$ are as follows:
\begin{subequations}
\begin{eqnarray}
\label{Eq:5a}
  U_x &=& x-\frac{1}{\Lambda}\Big\{\frac{\beta  x}{r_0^3}+\sum_{i=1}^{2}\tilde{x}_i\big(\frac{1}{r_i^3}+\frac{3A}{2r_i^5}\big)\Big\},\\
  \label{Eq:5b}
  U_y &=&y \Big[ 1-\frac{1}{\Lambda}\Big\{\frac{\beta  }{r_0^3}+\sum_{i=1}^{2}\big(\frac{1}{r_i^3}+\frac{3A}{2r_i^5}\big)\Big\}\Big],\\
  \label{Eq:5c}
 U_z &=&-\frac{z}{\Lambda}\Big\{\frac{\beta  }{r_0^3}+\sum_{i=1}^{2}\big(\frac{1}{r_i^3}+\frac{3A}{2r_i^5}\big)\Big\},\\
 \label{Eq:5d}
 \tilde{x}_i&=&x-x_i.
\end{eqnarray}
\end{subequations}
%The above dynamical system has two types of solutions, the libration points on the $xy-$plane and the libration points on the $xz-$plane.

From the Eq.(1), we observe that $\Lambda>0$, therefore the parameter $A$ must satisfy the condition $A>-\frac{(1+4\beta)}{3(1+8\beta)}$. In Fig. \ref{Fig:01N}, the evolution of the bifurcation curve are presented in the $(\beta, A)$ plane. In the area below the curve  there are non-permissible value of $(\beta, A)$, while  in the area above the curve there are permissible values of the combination of $(\beta, A)$. The red dashed line shows the permissible value of $A$ when $\beta=0$. It can be observed that for large value of $\beta$,  the value of $A$ becomes almost constant, i.e., for $\beta>10000$, $A=-0.166669$.
%%%%%%%%%%%%%%%%%%%%In plane libration points%%%%%%%%%%%%%
\subsection{In plane libration points}
\label{LP:01}
The in plane libration points are those libration points which lie on the $xy-$plane. When we consider the libration points on the $xy-$plane, i.e.,  $z=0$ and the third equation is always fulfilled. Therefore, the libration points can be found by solving the equations $U_x=0$, $U_y=0$ when $z=0$. There exist two type of libration points in the $xy-$plane, i.e., the collinear libration points and the non-collinear libration points.
\subsubsection{the collinear libration points}
\label{Sec:3a}
The collinear libration points are solutions of the equation \ref{Eq:5a}, when $y=z=0$,  i.e.,
\begin{equation}\label{Eq:3a:01}
  x-\frac{1}{\Lambda}\Big[\frac{\beta x}{|x|^3}+\frac{R^-}{|R^-|^3}+\frac{R^+}{|R^+|^3}+\frac{3A}{2}\Big\{\frac{R^-}{|R^-|^5}+\frac{R^+}{|R^+|^5}\Big\}\Big]=0
\end{equation}
where, $x-1/2= R^-$ and $x+1/2= R^+$.

 In this subsection, we wish to discuss the effect of oblateness or prolateness,  and the mass parameter on the positions of collinear libration points. According to \cite{DKM12}, $A <0$ corresponds to the case of prolate primaries whereas $A >0$ corresponds the case of oblate primaries and when $A=0$, the problem converts to the classical case.

 The parametric evolution of the positions of collinear libration points are illustrated in Fig. \ref{Fig:1N} for two different values of $\beta$. It is observed that the positions as well as the number of libration points on $x-$axis strongly depend on the combination of the values of $\beta$ and $A$. For $\beta=0.1$, there exist $8$ collinear libration points for $A\in( -0.039515, 0)$ whereas there exist only $4$ libration points for $A\in[0, 1)$ and $A\in[-0.259259, -0.039515)$.  The libration points $L^+_{x3, 4}$ and $L^-_{x3,4}$ originate from the vicinity of the primaries $m_1$ and $m_2$ respectively for slightly negative value of $A$, and $L^+_{x3}$, $L^-_{x3}$ coincide with $L^+_{x2}$,  $L^-_{x2}$ respectively and disappear completely at $A=- 0.039515$ whereas $L^+_{x4}$,  $L^-_{x4}$ exist for $A< -0.039515$ which we again labeled as $L^+_{x2}$, $L^-_{x2}$ for $A\in[-0.259259, -0.039515)$.    In addition, when $\beta=1000$, there exist $4$ collinear libration points for $A\in(-0.166687, -0.09805989) $ and  $A\in(0, 1)$ whereas there exist no collinear libration point when $A\in  (-0.09805989, 0)$.  In addition, we can notice that there originate a pair of collinear libration points in the vicinity of the each peripheral primaries,  i.e., there exist four collinear libration points in total.
\subsubsection{the non-collinear libration points}
\label{Sec:3b}
The non-collinear libration are solutions of Eqs (\ref{Eq:5a}, \ref{Eq:5b}) when $y\neq0$. In this subsection, we discuss the locations of the libration points evaluated numerically, by well known Newton-Raphson iterative scheme. This method  operates good concerning the convergence and the reliability of the results for the particular class of equations and this is  exactly the reason why we have used this method. It is necessary to note that the  Newton-Raphson method behaves well on the proper choice of the initial condition and consequently, for the good choice of initial conditions, we have used the  efficient and smart technique introduced by \cite{AAE16a}.

\begin{figure*}
\centering
\resizebox{\hsize}{!}{\includegraphics{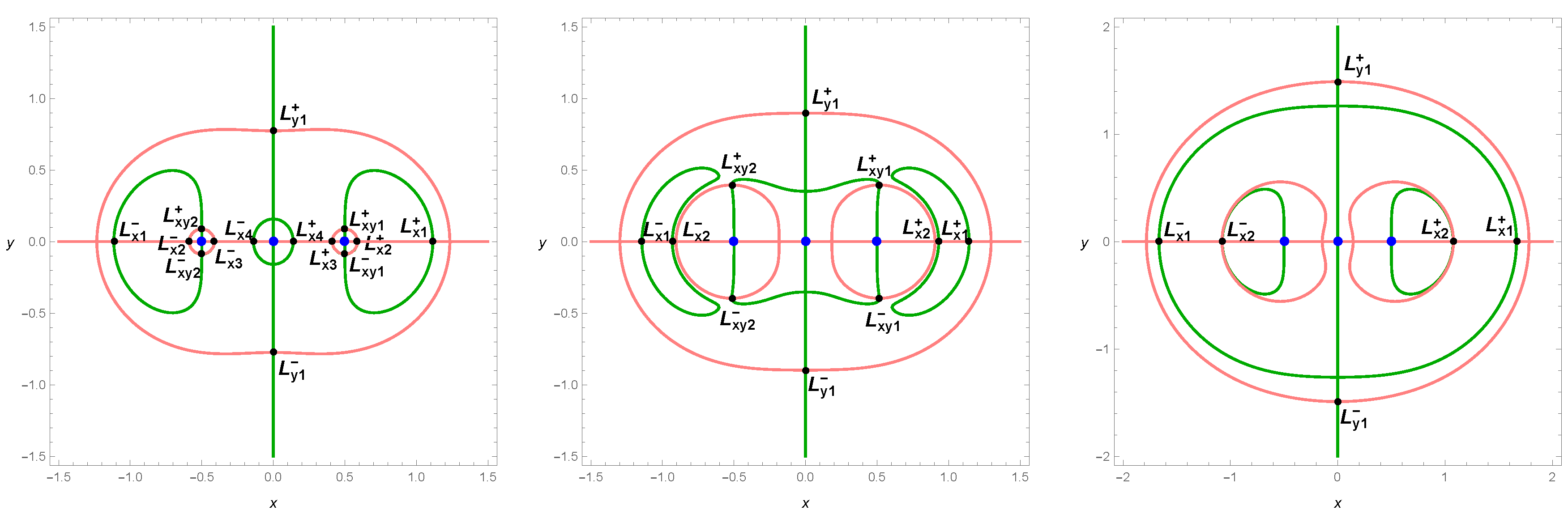}}
\resizebox{\hsize}{!}{\includegraphics{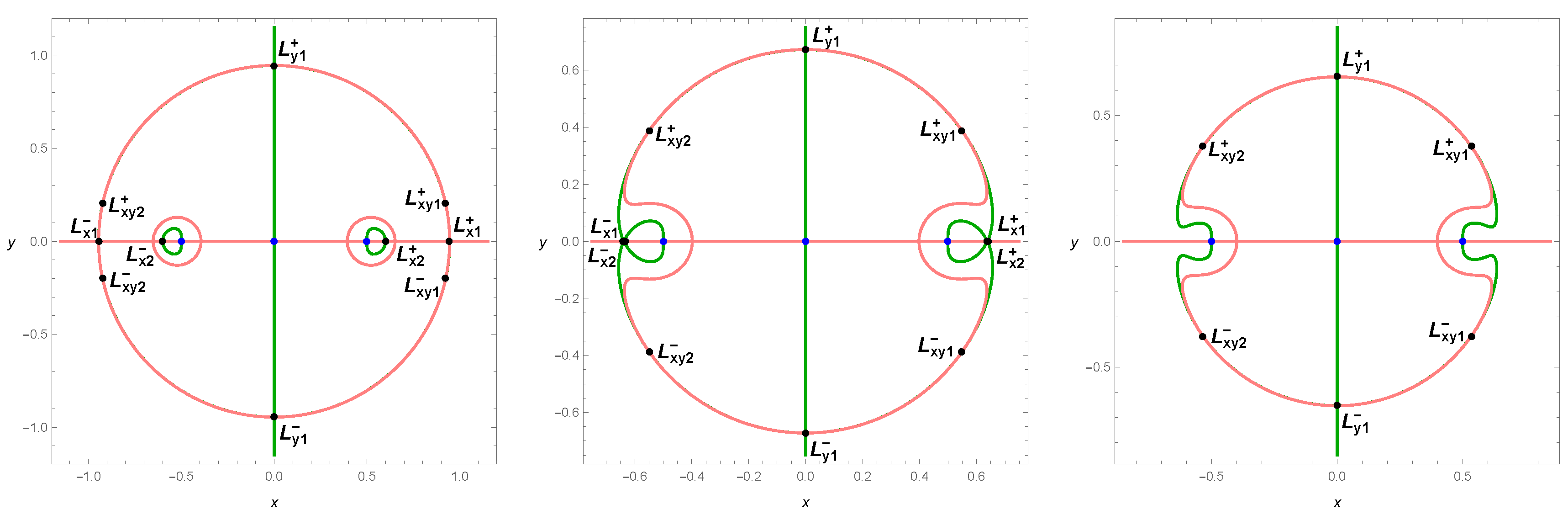}}
\caption{The evolution of the positions of the libration points,\emph{ first row}: for $\beta=0.1$ : (a) left: $ A=-0.005$; (b)middle: $ A=-0.1$; (c) right: $A=-0.215$:  \emph{second row}: for $\beta=1000$ : (d) left: $ A=-0.142$; (e)middle: $ A=-0.09804989$; (f) right: $  A=-0.092$.  The blue dots show the positions of primaries. (colour figure online).}
\label{Fig:04N}
\end{figure*}

In Fig. \ref{Fig:04N}, the parametric evolution of the positions of libration points on the $(x, y)$-plane are illustrated for the different values of $\beta$ and $A$. In Fig. \ref{Fig:04N}(a-c), the value of $\beta= 0.1$ is fixed and values of $A$ decreases. It is observed that there exist 14 libration points in which four libration points originate in the vicinity of the each of peripheral primaries. As the value of $A$ decreases, the
libration points $L^+_{x3}$, $L^+_{x4}$ and $L^-_{x3}$, $L^-_{x4}$ move towards each other and collide and finally disappear completely (see Fig. \ref{Fig:04N}b). Further decrease in the value of $A$ leads to the result that the libration points $L^+_{xy1,2}$ and $L^-_{xy1,2}$ disappear completely (see Fig. \ref{Fig:04N}c) and consequently there exist the libration points which lie only on the axes.

In Fig. \ref{Fig:04N}(d-e), we have illustrated the in-plane libration points for the value of $\beta=1000$ and observe completely different results regarding the existence and movement of the  libration points. This clearly indicates that the existence and the movement of the libration points depend on both the parameters, i.e., $\beta$ and $A$. It is observed that the libration points $L^+_{x1}$, $L^+_{x2}$ and $L^-_{x1}$, $L^-_{x2}$ move towards each other and disappear completely as the value of $A$ increases and consequently there exists no libration point on the $x$-axis  (see Fig. \ref{Fig:04N}f).

%%%%%%%%%%%%%%%%%%%%%%%%%%%%%%%%%%%%%%%%%%%%
%%%%%%%%%%%%%%%%%%%%%%%%%%%%%%%%%%%%%%%%%%%%%%%%%%
%%%%%%%%%%%%%%%%%%%%%%%%%%%%%%%%%%%%%%%%%%%%
\begin{figure}
\centering
\resizebox{\hsize}{!}{\includegraphics{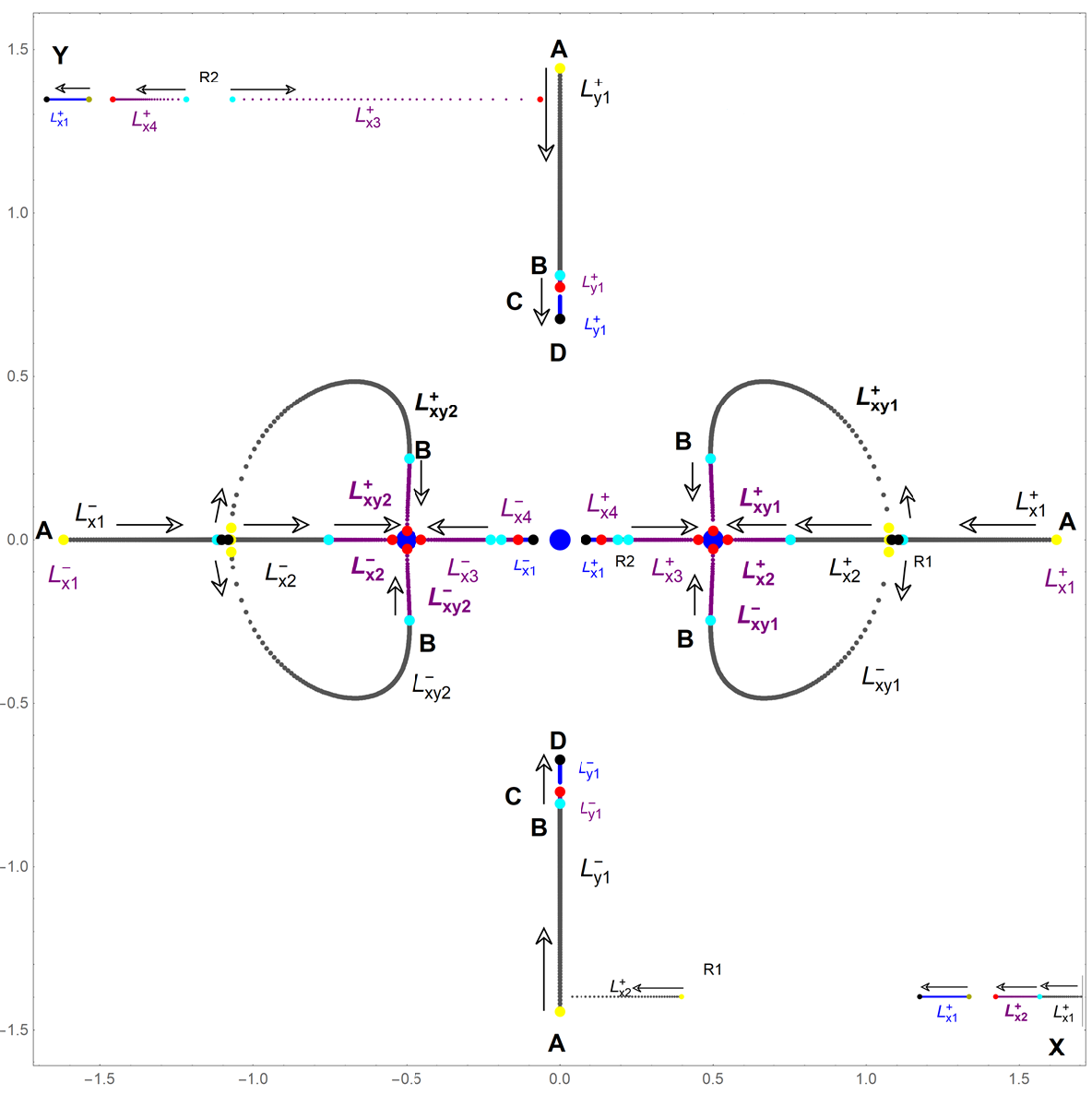}}
\caption{The movement of the positions of the libration points on $(x, y)$-plane  for $\beta=0.1$ and $A\in(-0.259259, 0.25)$,  the yellow dots show the value of the parameter $A=$\textbf{A} $\approx -0.259259$,  the cyan dots show the value of $A=\textbf{B}=-0.039515$, the red dots show the value of $A=$\textbf{C}$=0$, and  the black dots show the value of $A=$\textbf{D}$=0.25$. The arrow represents the movement of the positions of libration points while the big blue dots show the positions of primaries.  (colour figure online).}
\label{Fig:04NN1}
\end{figure}
%%%%%%%%%%%%%%%%%%%%%%%%%%%%%%%%%%%%%%%%%%%%%%%%%%
%%%%%%%%%%%%%%%%%%%%%%%%%%%%%%%%%%%%%%%%%%%%
\begin{figure}
\centering
\resizebox{\hsize}{!}{\includegraphics{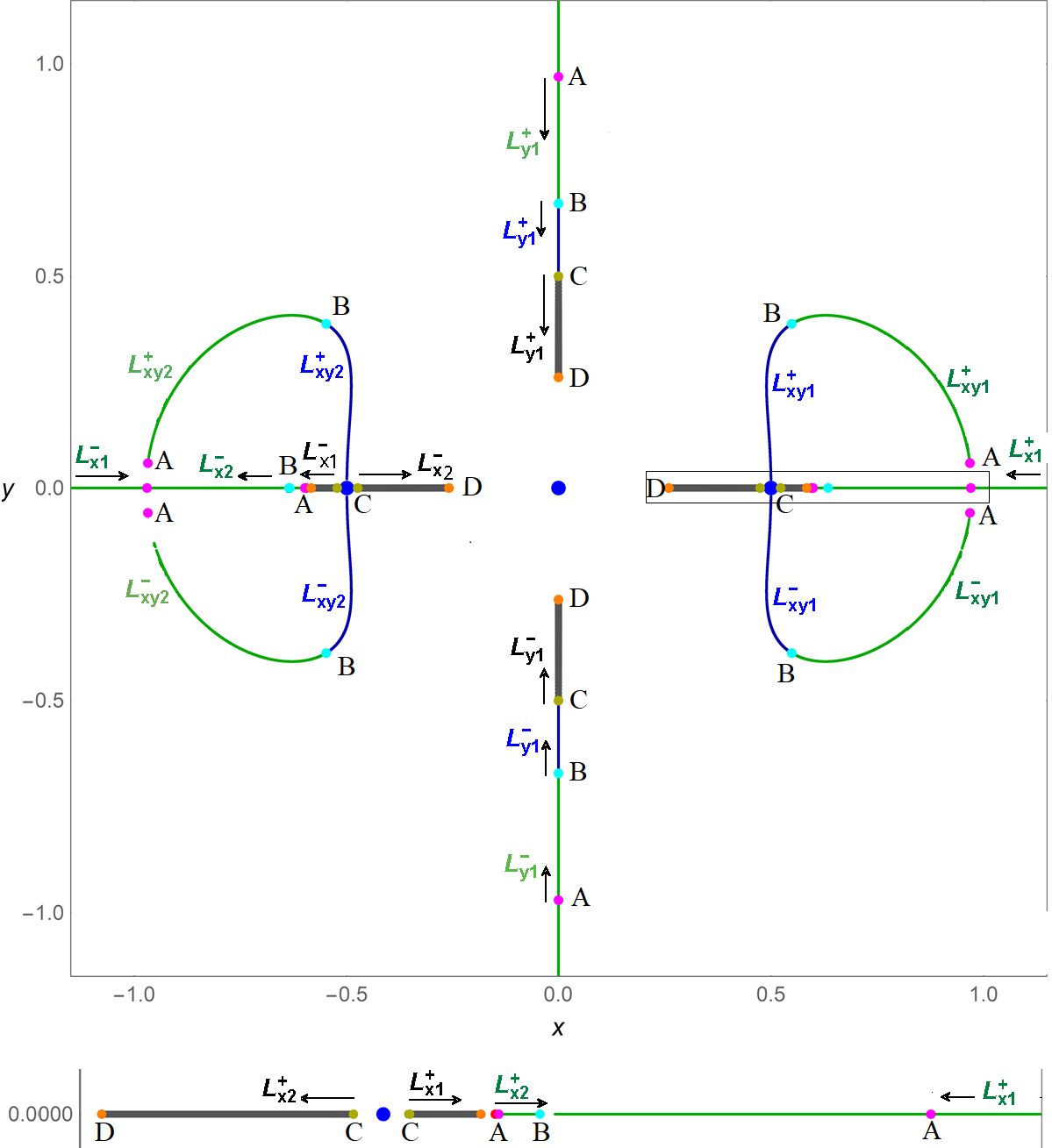}}
\caption{The movement of the positions of the libration points for $\beta=1000$ and the parameter $A\in(-0.166687, 1)$.  The \emph{magenta} dots show the value of parameter $A=$\textbf{A},  cyan dots show the value of $A=$\textbf{B}$=-0.097968$,  cyan dots show the value of $A=$\textbf{C}$= 0$, and orange dots show the value of $A=$\textbf{D} $\approx 1$.  The arrow shows the movement of the positions of libration points. The big blue dots show the positions of the primaries. (colour figure online).}
\label{Fig:04NN0}
\end{figure}
%%%%%%%%%%%%%%%%%%%%%%%%%%%%%%%%%%%%%%%%%%%%%%%%%%

In Fig.\ref{Fig:04NN1}, we have illustrated the parametric evolution of the positions of the libration points for varying value of the oblatenes/proletness parameter and fixed value of the mass parameter $\beta=0.1$. It can be observed that, for slightly higher value of the prolateness parameter $A$, there exist four collinear libration points whereas in the vicinity of libration points $L_{x2}^\pm$ there originate a pair of non-collinear  libration points $L_{xy1, xy2}^\pm$ which first move away from $x-$axis as the value of $A$ increases i.e., when $A\in (-0.259259, \textbf{B})$, four collinear libration points exist in which  two libration points lie on $y-$axis and four libration points lie on the  $xy$-plane.  The collinear libration points $L_{x1}^\pm$ and $L_{x2}^\pm$ move towards the peripheral primaries, the libration points $L_{y1}^\pm$ move towards the central primary whereas the non-collinear libration points $L_{xy1}^\pm$ and $L_{xy2}^\pm$ move far from the $x-$axis first then again turn towards the $x-$axis as $A\in (-0.259259, \textbf{B})$ and finally annihilate in the vicinity of the peripheral primaries as $A\in (\textbf{B, C})$.  Further, when $A\in(\textbf{B, C})$ we can observe that there exist eight collinear libration points in total and two new collinear  libration points $L_{x3}^\pm$ and  $L_{x4}^\pm$ originate in which the libration points $L_{x3}^\pm$ move towards peripheral primaries and at $A=0$  these libration points annihilate in the vicinity of them respectively, and  the libration points $L_{x4}^\pm$ move towards the central primary as the value of parameter $A$ increases and the movement of the remaining libration points are same. Further, when $A>0$, there exists no libration point on the $xy-$ plane, on the contrary the libration points always exist on either of the axes (shown in blue colour line). It can be observed that  the libration points $L_{x1}^\pm$ and $L_{y1}^\pm$ move towards the central primary $m_0$  along the $x-$axis and $y-$axis respectively.

In Fig. \ref{Fig:04NN0}, we have illustrated the parametric evolution of the positions of libration points for the fixed value of $\beta=1000$ and varying values of $A\in (-0.166687, 1)$.  The \emph{magenta, cyan, olive and orange} dots represented by \textbf{A, B, C,} and \textbf{D} respectively show the critical values of the oblateness/prolateness parameters where the number of libration points changes. It is observed that as the value of $A$  is slightly greater than the permissible value i.e., $-0.166687$,  six libration points exist in which four libration points namely $L_{x1, x2}^+$ and  $L_{x1, x2}^-$ exist on $x-$axis and two libration points $L_{y1}^+$ , $L_{y1}^-$ exist on $y-$axis. As the value of $A$ approaches to  \textbf{A} (shown by magenta dots in  Fig. \ref{Fig:04NN0}), four non-collinear libration points namely $L_{xy1}^+, L_{xy1}^-$ and $L_{xy2}^+, L_{xy2}^-$ originate in  the vicinity of the collinear libration points $L_{x1}^+$ and $L_{x1}^-$ respectively, and it can be further observed that these libration points move away from the $x-$axis whereas the collinear libration points $L_{x1}^{\pm}$ move towards  and  $L_{x2}^\pm$ move away from the peripheral primaries as value of $A$ increases. In addition, at \textbf{B} (shown by cyan dot) the collinear libration points $L_{x1}^{\pm}$ collide with $L_{x2}^{\pm}$ respectively, and disappear. Consequently, there exists no collinear libration point for prolateness  parameter $A \in$ (\textbf{B}, \textbf{C}) (shown in darker blue line) and only the non-collinear libration points $L_{xy1,2}^\pm$ and $L_{y1}^\pm$ exist in this range and at \textbf{C}, the non-collinear libration points $L_{xy1,2}^\pm$ collide with the peripheral primaries and disappear completely. Further, when $A\in($\textbf{C}$, 0.25)$, a pair of collinear  libration points originate in the vicinity of each of the peripheral primaries and move far from them as the value of $A$ increases (shown in darker gray line). In this interval, for the values of the oblateness parameter,  there exist six libration points in total which lie on either of the axes (shown in darker gray line). It is noticed that as the value of oblateness/prolateness parameter $A$ increases the libration points $L_{y1}^\pm$ always move towards the origin along $y-$axis.
%%%%%%%%%%%%%%%%%%%%%%%%%%%%%%%%%%%%%%%%%%%%%%%%%%%
%%%%%%%%%%%%%%%%%%%%%%%%%%%%%%%%%%%%%%%%%%%%
\begin{figure*}
\centering
\resizebox{\hsize}{!}{\includegraphics{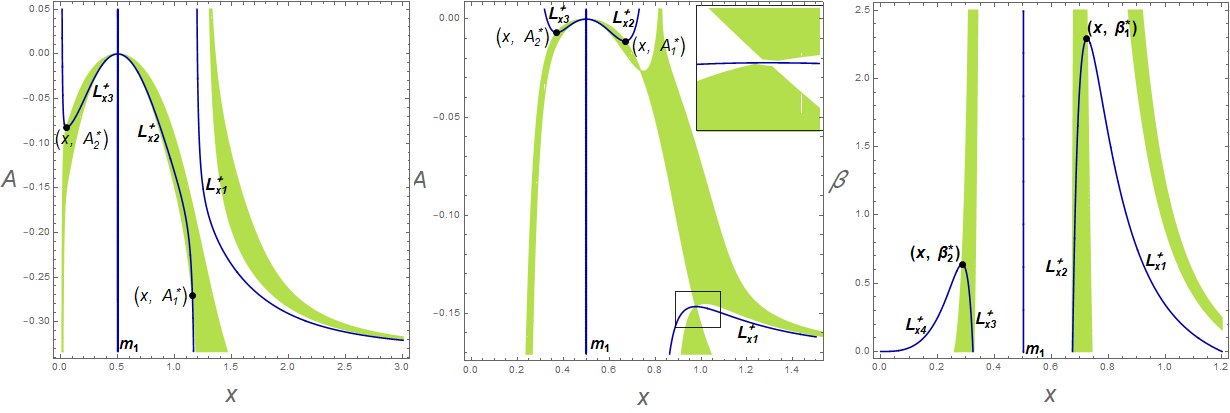}}
\caption{The parametric evolution of the stability of collinear libration points: (a) \emph{left}:  for $\beta=0.0001$ and $(x, A_1^*)=(1.15454220, -0.27055522)$, and $(x, A_2^*)=(0.05161635, -0.08262583)$;  (b) \emph{middle}:  for $\beta=5$ and $(x, A_1^*)=(0.66910678, -0.01123429)$, and $(x, A_2^*)=(0.37141443, -0.00690825)$;  (c) \emph{right}:  for $A=-0.02$ and $(x, \beta_1^*)=(0.72564641, 2.29207325)$, and $(x, \beta_2^*)=(0.28636778, 0.63939639)$.   The stable regions are depicted by \emph{green} colour.  (colour figure online).}
\label{Fig:CSTAB1}
\end{figure*}
%%%%%%%%%%%%%%%%%%%%%%%%%%%%%%%%%%%%%%%%%%%%
\begin{figure*}
\centering
\includegraphics[width=\columnwidth]{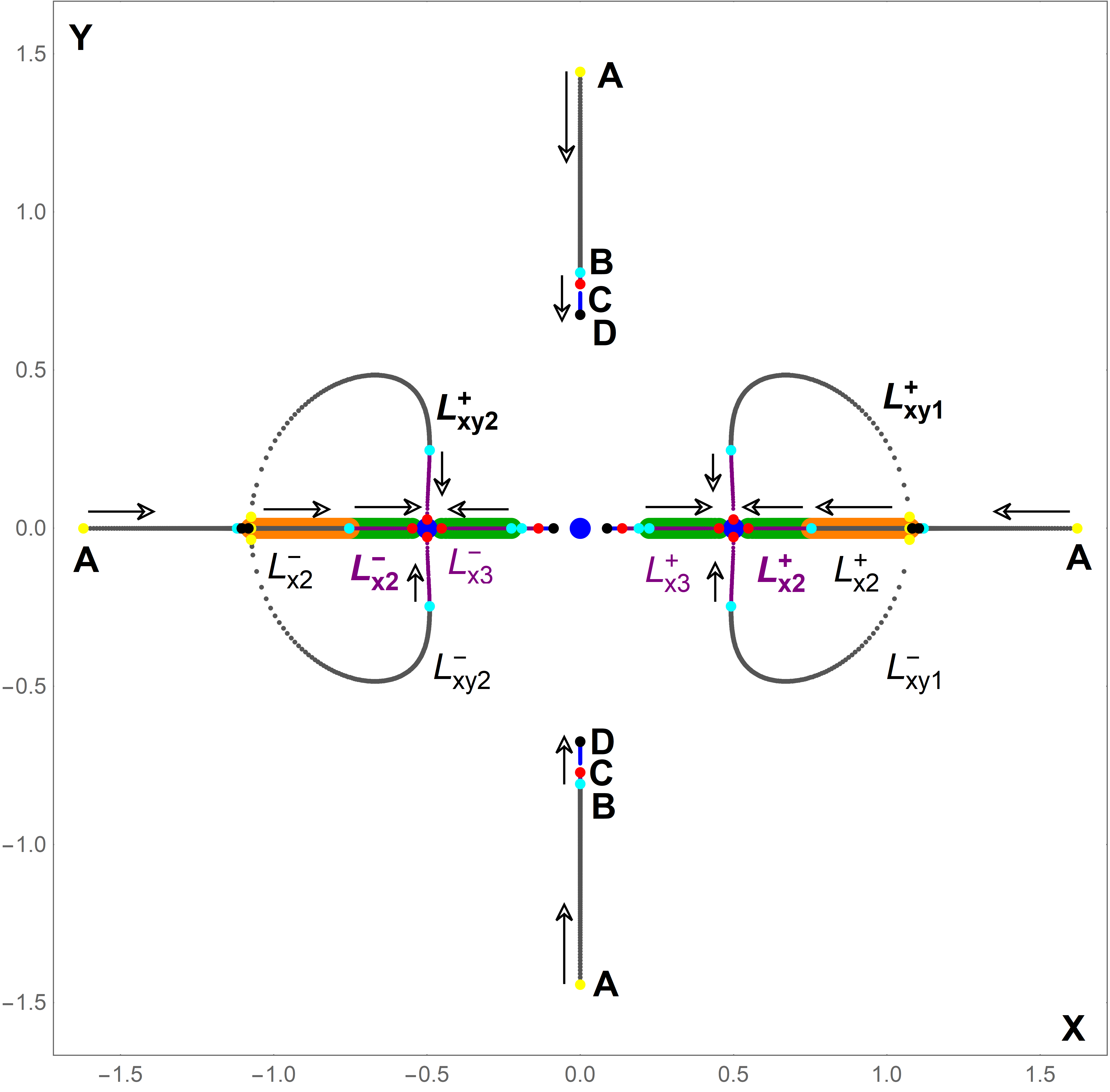}
\includegraphics[width=\columnwidth]{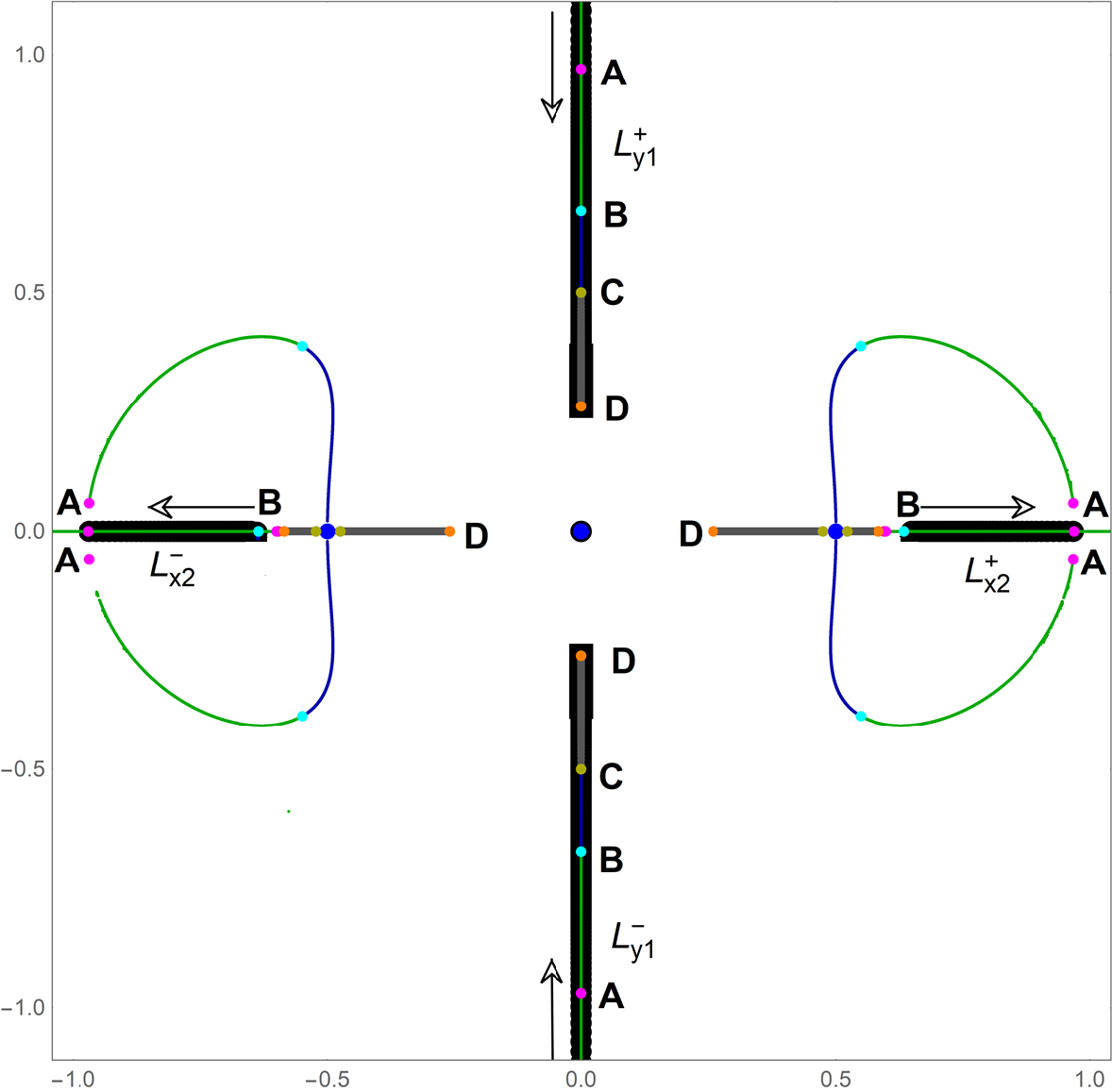}\\
\caption{The parametric evolution of the stability regions of the libration points for  (a) \emph{left}: $\beta=0.1$  and $A\in(-0.259259, 0.25)$, here the thick orange and green lines show the positions of the stable libration points. The value of \textbf{A, B, C}, and  \textbf{D} are same as in Fig. \ref{Fig:04NN1}; (b) \emph{right}: $\beta=1000$ and $A\in(-0.166687, 1)$, thick black lines show the positions of stable libation points. The values of \textbf{A, B, C}, and  \textbf{D} are same as in Fig. \ref{Fig:04NN0}.  (colour figure online).}
\label{Fig:STABXY04NN2}
\end{figure*}
%%%%%%%%%%%%%%%%%%%%%%%%%%%%%%%%%%%%%%%%%%%%%%%%%%
%%%%%%%%%%%%%%%%%%%%%%%%%%%%%%%%%%%%%%%%%%%%
\begin{figure*}
\centering
\resizebox{\hsize}{!}{\includegraphics{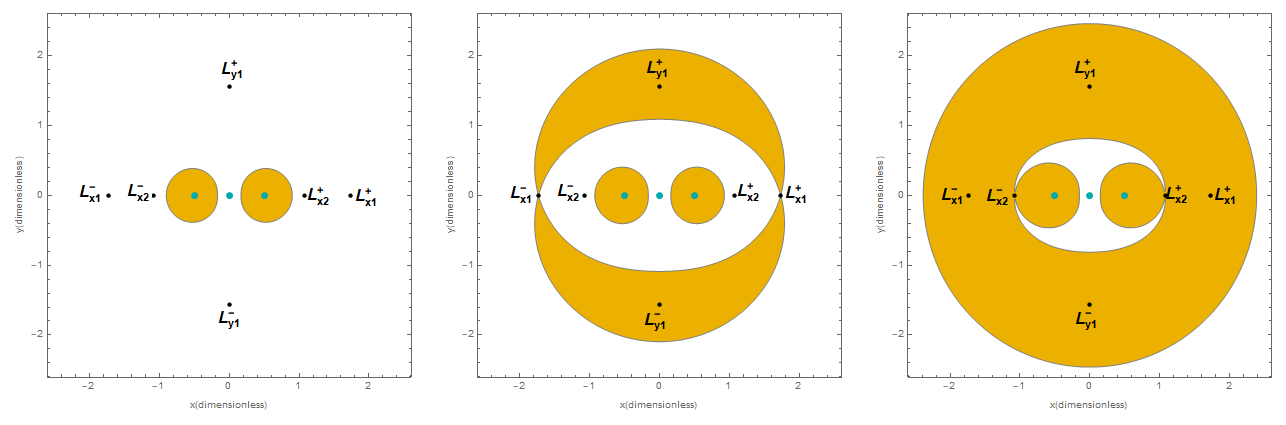}}\\
\resizebox{\hsize}{!}{\includegraphics{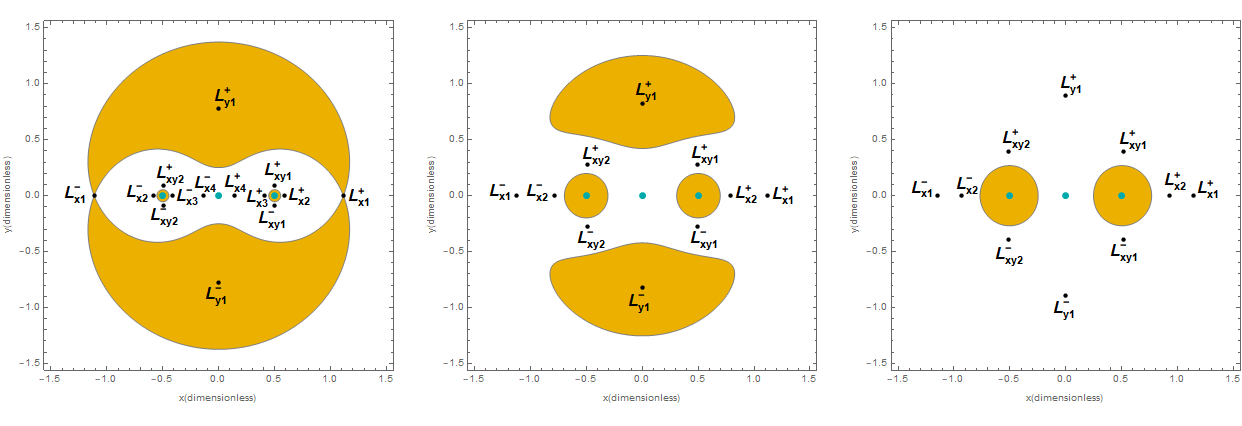}}
\caption{The evolution of the regions of possible motion:\emph{ first row}: for $\beta=0.1, A=-0.22$: (a)left: $C=8.25875064$; (b) middle: $C=8.89008187$; (c)right: $C=9.93069941$:, \emph{second  row}: for $C=2.93554384$, (a) left: $A=-0.005$, (b) middle: $A=-0.05$, (c) right: $A=-0.1$.  (colour figure online).}
\label{Fig:06N}
\end{figure*}
%%%%%%%%%%%%%%%%%%%%%%%%%%%%%%%%%%%%%%%%%%%%
%%%%%%%%%%%%%%%%%%%%%The Stability of Libration points%%%%%%%%%%%%%%%%%
\section{The Stability of libration points}
\label{S:0}
In this section, we deal with the stability of the libration points in the configuration $(x, y)$-plane, where the effects of the parameters $A$ and $\beta$ on the stability of these libration points are illustrated.
%%%%%%%%%%%%%%%%%%%%%%%%%%%%%%%%%%%%%%%%%%%%%%%%%
%\subsection{stability of the in-plane libration points}
The stability of the libration points can be determined by linearizing the equations of motion of the test particle given in Eqs. \ref{Eq:1a}-\ref{Eq:1b} about the libration point $(x_*, y_*)$.  Consequently, the linearized equations for the test particle near the  libration points in the collinear restricted four-body problem are $\dot{\mathbf{x}}=\mathbb{A}\mathbf{x}$, where $\mathbf{x}=(x, y, \dot{x}, \dot{y})^T$, and $\mathbf{x}$ is the state vector of the infinitesimal body with respect to the libration points and $A$, the coefficient matrix is read as:
\begin{eqnarray}
\label{Eq:16}
\mathbb{A} &=& \left(
                    \begin{array}{cc}
                     \mathbb{ O} & \mathbb{I } \\
                     \mathbb{ B} & \mathbb{C}   \\
                     %A_{31}&  A_{32} &  A_{33} &  A_{34} \\
                     %A_{41} &  A_{42}&  A_{43}&  A_{44} \\
                    \end{array}
                  \right),
\end{eqnarray}
where
\begin{eqnarray*}
       \mathbb{O} &=&\left(
                    \begin{array}{cc}
                     0 & 0\\
                     0 & 0  \\
                    \end{array}
                    \right),
                     \mathbb{I}=\left(
                    \begin{array}{cc}
                     1 & 0\\
                     0 &1  \\
                    \end{array}
                    \right),\\
                    \mathbb{B}& =&\left(
                    \begin{array}{cc}
                     A_{11}&  A_{12} \\
                     A_{21} &  A_{22} \\
                    \end{array}
                    \right),
                    %%%%%%%%%%%%%%%%%%%%%%%%%
                    \mathbb{C}=\left(
                    \begin{array}{cc}
                    0&   2 \\
                    -2 &  0\\
                    \end{array}
                    \right).
     \end{eqnarray*}
Accordingly, the characteristic equation corresponding to the matrix given in Eq. \ref{Eq:16} is
\begin{equation}\label{Eq:17}
 \lambda^4+a_1 \lambda^2+a_0=0,
\end{equation}
where
\begin{eqnarray*}
  a_1 &=&4-A_{11}-A_{22},\\
  a_0 &=&A_{11}A_{22}-A_{12}^2,\\
  %\end{eqnarray*}
%  \begin{eqnarray*}
  A_{11}&=&1-\frac{1}{\Lambda}\Big(\frac{\beta}{r_{0}^{3}}+\sum_{i=1}^{2}\Big( \frac{1}{r_{i}^{3}}+\frac{3 A}{2r_{i}^5} \Big)\Big)+\frac{1}{\Lambda} \Big( \frac{3\beta x_*^2}{r_0^5} \\
  &&+\sum_{i=1}^{2}\tilde{x}_i^2 \Big(\frac{3}{r_i^5}+\frac{15A}{2r_i^7} \Big) \Big),
  \end{eqnarray*}
\begin{eqnarray*}
  A_{22}&=&1-\frac{1}{\Lambda} \Big(\frac{\beta}{r_{0}^{3}}+\sum_{i=1}^{2}\Big( \frac{1}{r_{i}^{3}}+\frac{3 A}{2r_{i}^5} \Big)\Big)+\frac{y_*^2}{\Lambda} \Big( \frac{3\beta }{r_0^5} \\
  &&+\sum_{i=1}^{2} \Big(\frac{3}{r_i^5}+\frac{15A}{2r_i^7} \Big) \Big),\\
  A_{12}&=&\frac{y_*}{\Lambda}\Big(\frac{3\beta x_*}{r_0^5}+\sum_{i=1}^{2}\tilde{x_i} \Big( \frac{3}{r_i^5}+\frac{15A}{2r_i^7} \Big) \Big),\\
  &=&  A_{21}.\\
%\end{eqnarray*}
%\begin{equation*}\label{Eq:}
\tilde{x}_i&=&x_*-x_i,\\
  r_i&=&\sqrt{(x_*-x_i)^2+y_0^2}, i=0,1,2.
\end{eqnarray*}
A libration point is said to be stable if the solution of the Eq.\ref{Eq:17},  evaluated at the libration point, has four pure imaginary roots. This is true only when the conditions,
\begin{equation}
  a_1^2-4a_0 > 0, \quad
  a_1> 0,  \quad  a_0 > 0,
\end{equation}
are satisfied simultaneously.

In Fig. \ref{Fig:CSTAB1}, we have illustrated the linear stability of the collinear equilibrium points. In Fig. \ref{Fig:CSTAB1}a and Fig. \ref{Fig:CSTAB1}b, the stability of the equilibrium points are depicted for constant value of $\beta=0.0001$ and $\beta=5$ respectively. In both the cases, we observed that equilibrium points $L_{x2}^{\pm}$ and $L_{x3}^{\pm}$ are stable. However, as the value of $\beta$ increases,  the intervals of the values of parameter $A$ decreases in which the libration points are stable. In Fig. \ref{Fig:CSTAB1}c, the stability of collinear libration points are presented for fixed value of $A=-0.02$ and varying values of $\beta$ and observed that $L_{x2}^{\pm}$  and $L_{x3}^{\pm}$ are stable for $\beta\in(0, 2.29207325)$ and $\beta\in(0, 0.63939639)$ respectively.

In Fig. \ref{Fig:STABXY04NN2}, the stability of the libration points for $\beta=0.1$ and  $1000$ is presented for varying values of the parameter $A$. In Fig. \ref{Fig:STABXY04NN2}a, the stability of the libration points is presented for mass parameter $\beta=0.1$.  It is unveiled that the libration points $L_{x2}^{\pm}$ are linearly stable for $A\in ( \textbf{A}, \textbf{B}) $ and $A\in(\textbf{B, C})$ (the stable equilibrium points are depicted in thick \emph{orange} and \emph{green} lines respectively).  Further, the equilibrium points $L_{x3}^\pm$ are linearly stable for $A\in (\textbf{B,  C})$.

 In Fig. \ref{Fig:STABXY04NN2}b, the stability of the libration points is presented for $\beta=1000$ and we noticed that the collinear libration points $L_{x2}^\pm$ are stable for $A\in (\textbf{A, B})$ whereas $L_{y1}^\pm$ are also stable for $A\in(\textbf{A, B})$.  In addition, we have observed that the $L_{y1}^\pm$ are stable for the value of $A\in(\textbf{B, C})$ and $A\in(\textbf{C}, 1)$ also.  It is also observed that none of the libration points which lie on $(x, y)$ plane are stable, however, for some mass ratio the equilibrium  points which lie on $y-$ axis are linearly stable.
 %%%%%%%%%%%%%%%%%%%%%%%%%%%%%%%%%%%%%%%%%%%%%%%%%%
%%%%%%%%%%%%%%%%%%%%%%%%%%%%%%%%%%%%%%%%%%%%%%%%
\section{The regions of possible motion}
\label{ZVC:0}
 By using the relation \ref{Eq:3N}, we shall draw the evolution of the zero-velocity curves for fixed value of $C$, by assuming the motions of the particle  on the $xy$-plane. The zero-velocity curves bifurcate the regions of possible motion of those planes from the forbidden region where  the test particle can not orbit.
%%%%%%%%%%%%%%%%%%%%%%%%%%%%%%%%%%%%%%%%%%%%
\begin{figure*}
\centering
\resizebox{\hsize}{!}{\includegraphics{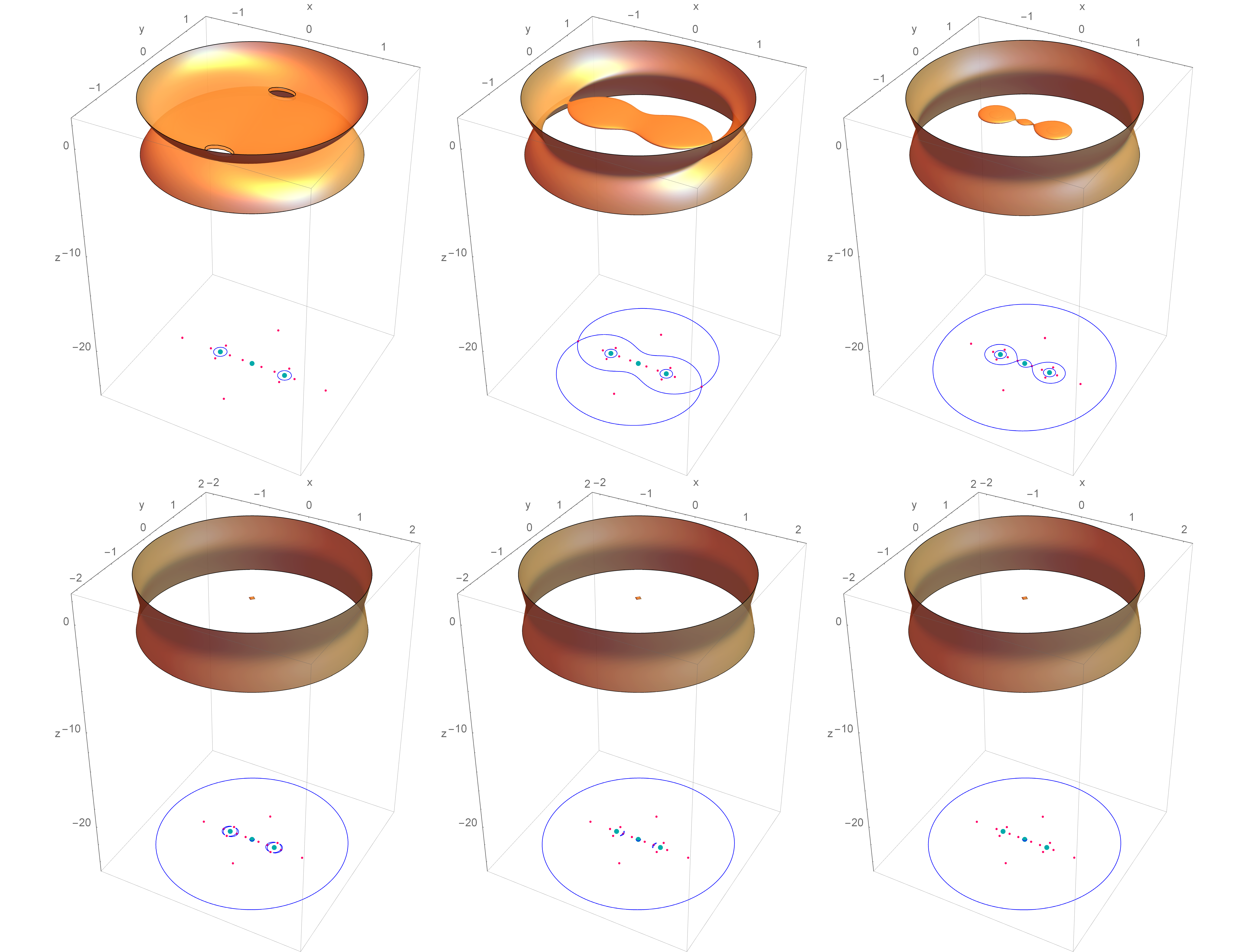}}
\caption{The evolution of the regions of possible motion when $A= -0.015$; $\beta= 0.1$: (a) top left: $C=2.32787248$; (b) top middle: $C=2.98763159$; (c) top right: $C= 3.70660029$; (d) bottom left: $C= 4.53152976$; (e) bottom middle: $C=4.56457186$;  (f) bottom tight: $C=4.59237937$.   (colour figure online).}
\label{Fig:08N}
\end{figure*}
%%%%%%%%%%%%%%%%%%%%%%%%%%%%%%%%%%%%%%%%%%%%%%%%%%

In Fig. \ref{Fig:06N}(a, b, c), the regions of possible motion are depicted for the fixed value of $\beta$, the prolateness parameter $A$ and increasing value of the Jacobian constant $C$. The coloured area represents the forbidden regions where the test particle can not communicate. At $C= C_{L^+_{y1}}=C_{L^-_{y1}}$, there exist two circular islands containing each of the peripheral primaries where the motion of the test particle is not possible. Therefore, the test particle can orbit from central primary to any libration points whereas it can not orbit to peripheral primaries. In \ref{Fig:06N}b, when the Jacobian constant $C$ is increased to $C= C_{L^+_{x1}}=C_{L^-_{x1}}$ the forbidden region increased and two crescent shapes appear which originate from $L^+_{x1}$ or $L^-_{x1}$ and contain the libration points $L^+_{y1}$ or $L^-_{y1}$, respectively. Consequently, the infinitesimal mass cannot approach to these libration points. Further increase in value of Jacobian constant to $C= C_{L^+_{x2}}=C_{L^-_{x2}}$  leads to further increase in the forbidden regions and consequently all the libration points fall inside the forbidden region. The test particle can move either outside the circular annulus shaped region or in the vicinity of the central primary.  Therefore, we can conjuncture that as the value of Jacobian constant increases, the forbidden region also increases.

In Fig. \ref{Fig:06N}(d, e, f), the ZVCs are depicted  for the fixed values of $\beta, C$ and the varying values of prolateness parameter $A$. It can be noticed that when $A=-0.005$, there exist four branches of forbidden regions, in which two are  circular in shaped which contain the peripheral primaries whereas two crescent shaped regions contain the libration points which exist on $y-$axis. However, the motion of the infinitesimal mass is possible inside the pear shaped region  which contains all  the libration points except those which lie on the $y-$ axis. As we decrease the value of parameter $A=-0.05$, the crescent shape regions containing libration points $L^+_{y1}$ and $L^-_{y1}$ shrink and further shrink to libration points $L^+_{y1}$ and $L^-_{y1}$ as the value of $A$ approaches to $\approx-0.08741$. It is observed that the circular shaped forbidden regions containing the peripheral primaries do exist  even for higher values of $A$. Consequently,  the test particle can not communicate from peripheral primaries to any other and vice-versa.

In Fig. \ref{Fig:08N}(a-f), the regions of possible motion are depicted for fixed value of the parameters $A$ and $\beta$ and increasing values of the Jacobian constant.  It is observed that the regions of possible motion decrease significantly when the  value of the Jacobian constant increases.
%%%%%%%%%%%%%%%%%%%%%%%%%%%%%%%%%%%%%%%%%%%%%
%%%%%%%%%%%%%%%%%%%%%%%%%%%%%%%%%%%%%%%%%%%%
%%%%%%%%%%%%%%%%%%%%%%%%%%%%%%%%%%%%%%%%%%%%
\begin{figure*}
\centering
\resizebox{\hsize}{!}{\includegraphics{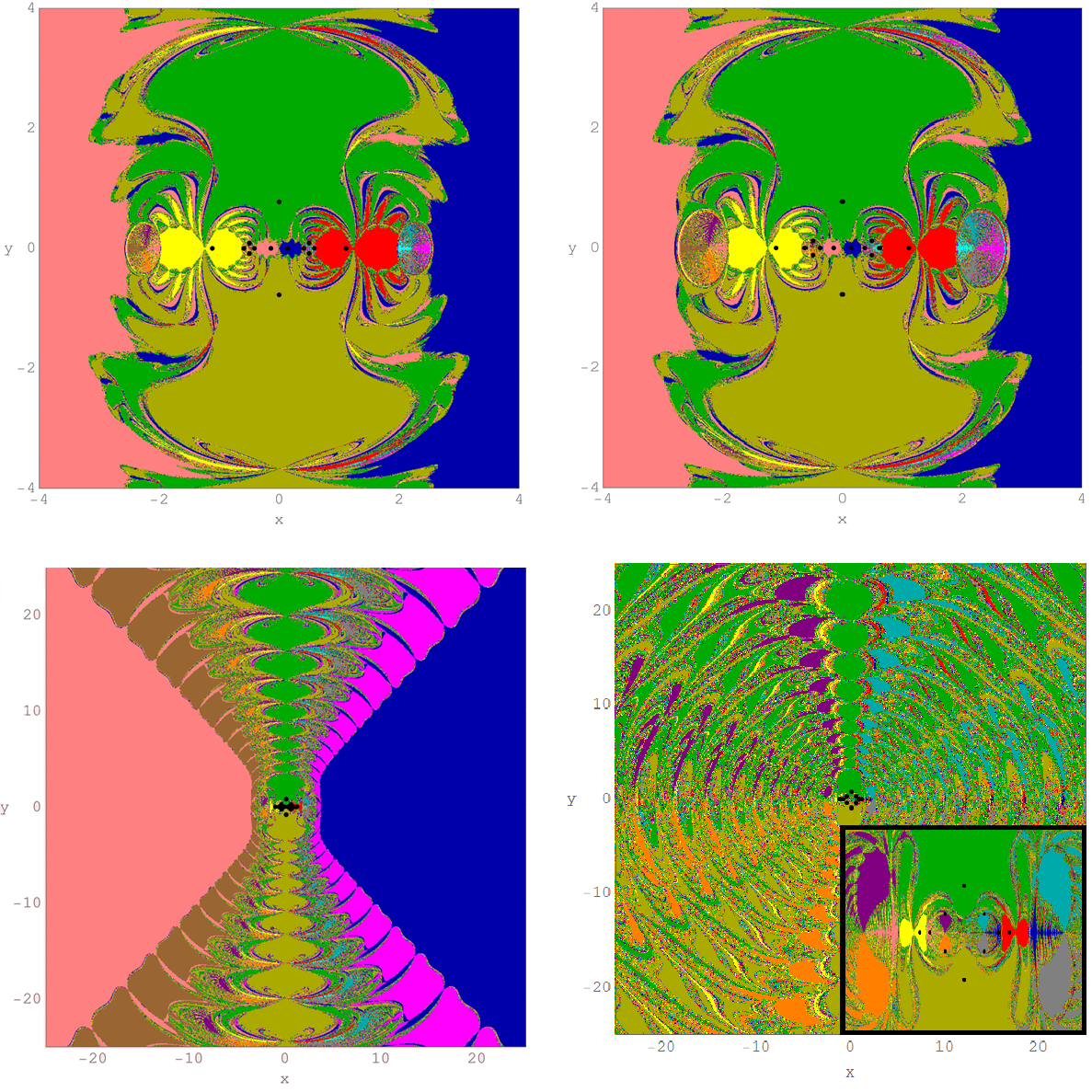}}
\caption{The BoC linked with the libration points on $(x, y)$-plane  for  $\beta=0.1$, and for  (a) $A=-0.005$, (b) $A=-0.01$, (c) $A=-0.0391$, (d) $A=-0.075$. The color codes for the BoC are as follows: $L_{x1}^+=$\emph{red}, $L_{x2}^+=$\emph{cyan}, $L_{x3}^+=$\emph{magenta}, $L_{x4}^+=$\emph{blue},  $L_{x1}^-=$\emph{yellow}, $L_{x2}^-=$\emph{light pink}, $L_{x3}^-=$\emph{brown}, $L_{x4}^-=$\emph{pink}, $L_{y1}^+=$\emph{green}, $L_{y1}^-=$\emph{olive},
$L_{xy1}^+=$\emph{teal},  $L_{xy2}^{+}=$\emph{purple}, $L_{xy1}^-=$\emph{gray} and $L_{xy2}^{-}=$\emph{orange}.
The dots show the  positions of libration points.  (colour figure online).}
\label{Fig:Basin_0}
\end{figure*}
%%%%%%%%%%%%%%%%%%%%%%%%%%%%%%%%%%%%%%%%%%%%
%%%%%%%%%%%%%%%%%%%%%%%%%%%%%%%%%%%%%%%%%%%%
%%%%%%%%%%%%%%%%%%%%%%%%%%%%%%%%%%%%%%%%%%%%
\begin{figure*}%
\centering
\includegraphics[width=\columnwidth]{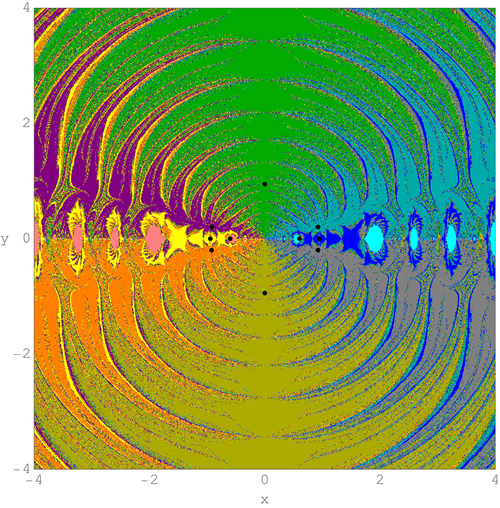}
\includegraphics[width=\columnwidth]{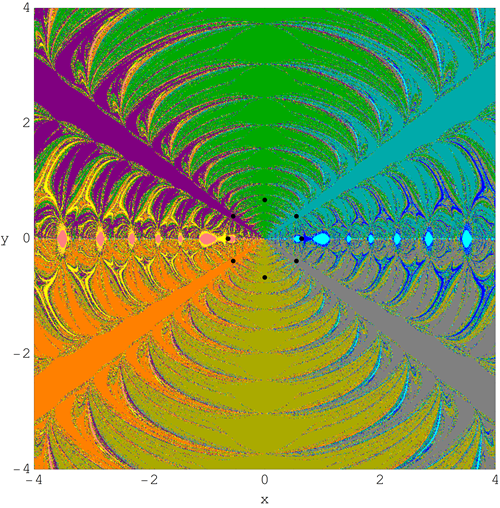}\\
\caption{The BoC linked with the libration points on $(x, y)$-plane  for  $\beta=1000$, and  (a) for $A=-0.142$, (b) $A=-0.09804989$. The color codes for the BoC are as follows: $L_{x1}^+=$\emph{blue}, $L_{x2}^+=$\emph{cyan},
 $L_{x1}^-=$\emph{yellow}, $L_{x2}^-=$\emph{pink}, $L_{y1}^+=$\emph{green} and $L_{y1}^-=$\emph{olive},  $L_{xy1}^+=$\emph{teal}, $L_{xy2}^+=$\emph{purple}, $L_{xy1}^-=$\emph{gray}, $L_{xy2}^-=$\emph{orange}.
The dots show the  positions of libration points.  (colour figure online).}
\label{Fig:Basin_2}
\end{figure*}
%%%%%%%%%%%%%%%%%%%%%%%%%%%%%%%%%%%%%%%%%%%%%%%%%%
\begin{figure*}
\centering
\includegraphics[width=\columnwidth]{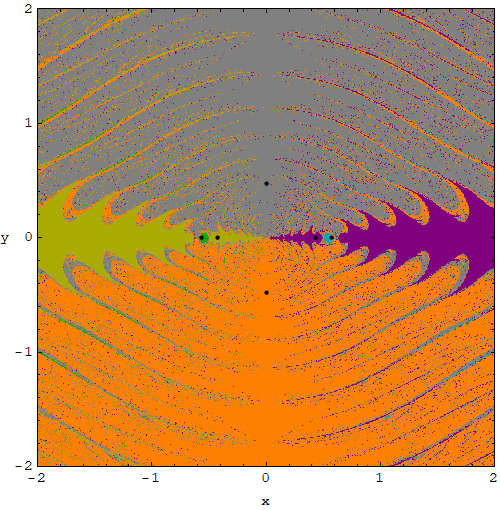}
\includegraphics[width=\columnwidth]{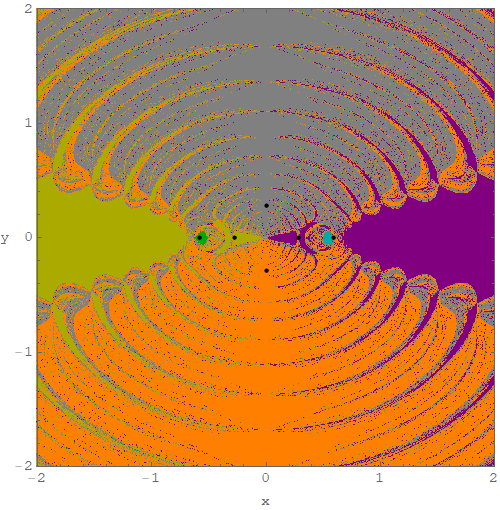}\\
\includegraphics[width=\columnwidth]{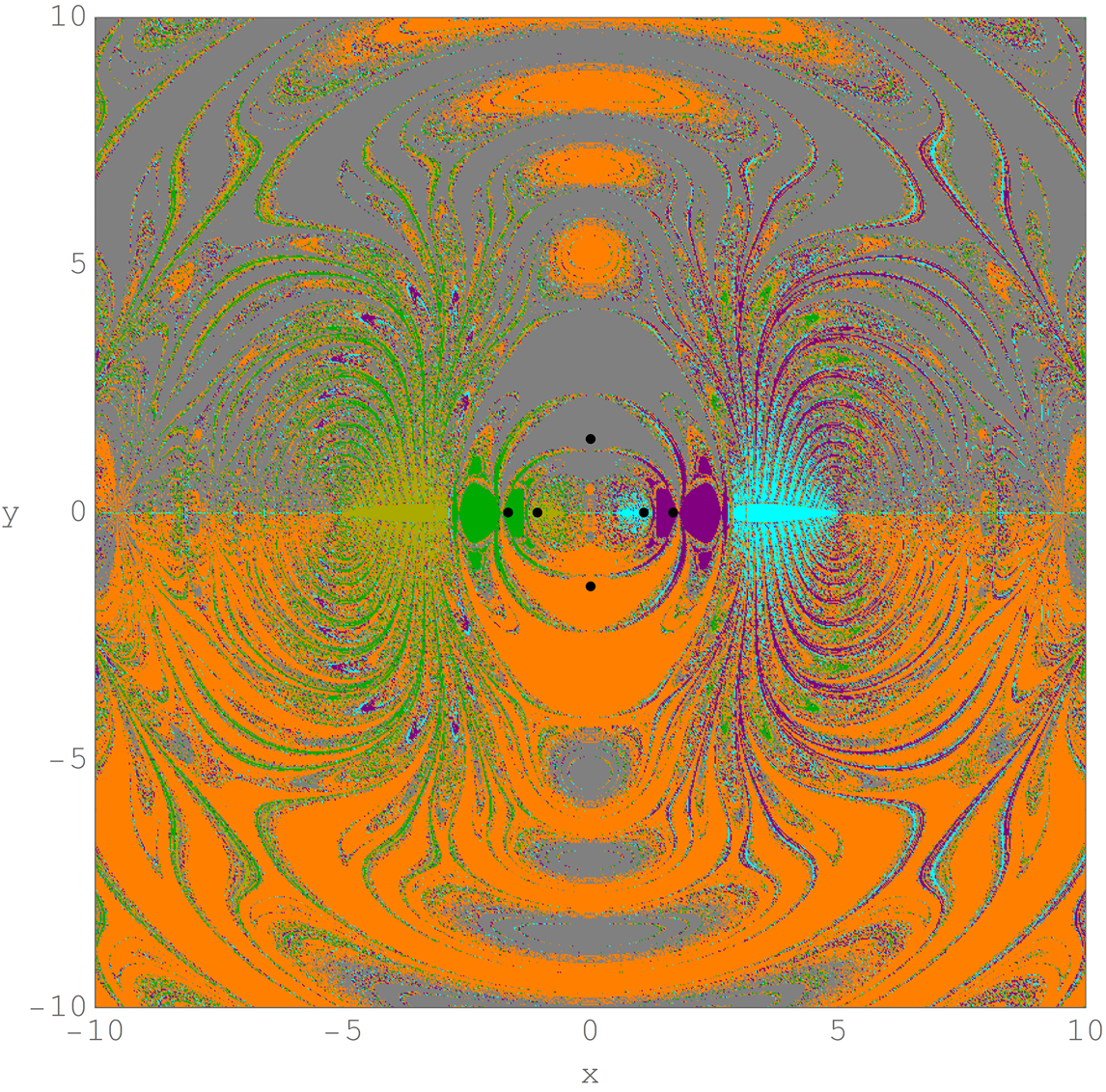}\\
\caption{The BoC linked with the libration points on $(x, y)$-plane  for  $\beta=1000$, and  (a) for $A=0.025$, (b) $A=0.75$. For  $\beta=0.1$, and  (c) for $A=-0.215$. The color codes for the BoC are as follows: $L_{x1}^+=$\emph{purple}, $L_{x2}^+=$\emph{cyan}, $L_{x1}^-=$\emph{green}, $L_{x2}^-=$\emph{olive}, $L_{y1}^+=$\emph{gray} and $L_{y1}^{-}=$\emph{orange}.
The dots show the  positions of libration points.  (colour figure online).}
\label{Fig:Basin_1}
\end{figure*}
%%%%%%%%%%%%%%%%%%%%%%%%%%%%%%%%%%%%%%%%%%%%%%%%%%
%%%%%%%%%%%%%%%%%%%%%%%%%%%%%%%%%%%%%%%%%%%%%%%%%%
%%%%%%%%%%%%%%%%%%%%%%%%%%%%%%%%%%%%%%%%%%%%
\begin{figure*}%
\centering
\includegraphics[width=\columnwidth]{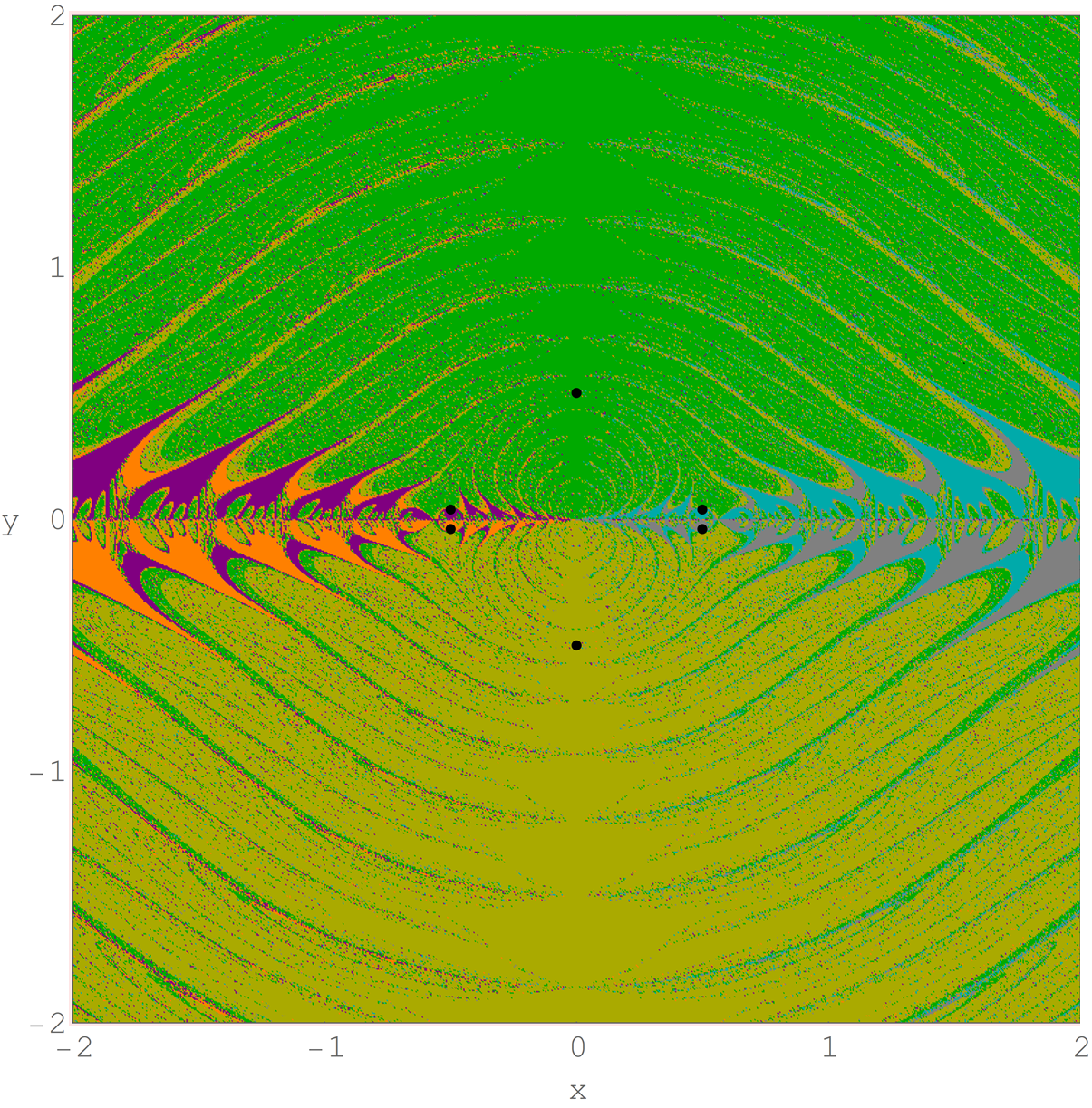}
\includegraphics[width=\columnwidth]{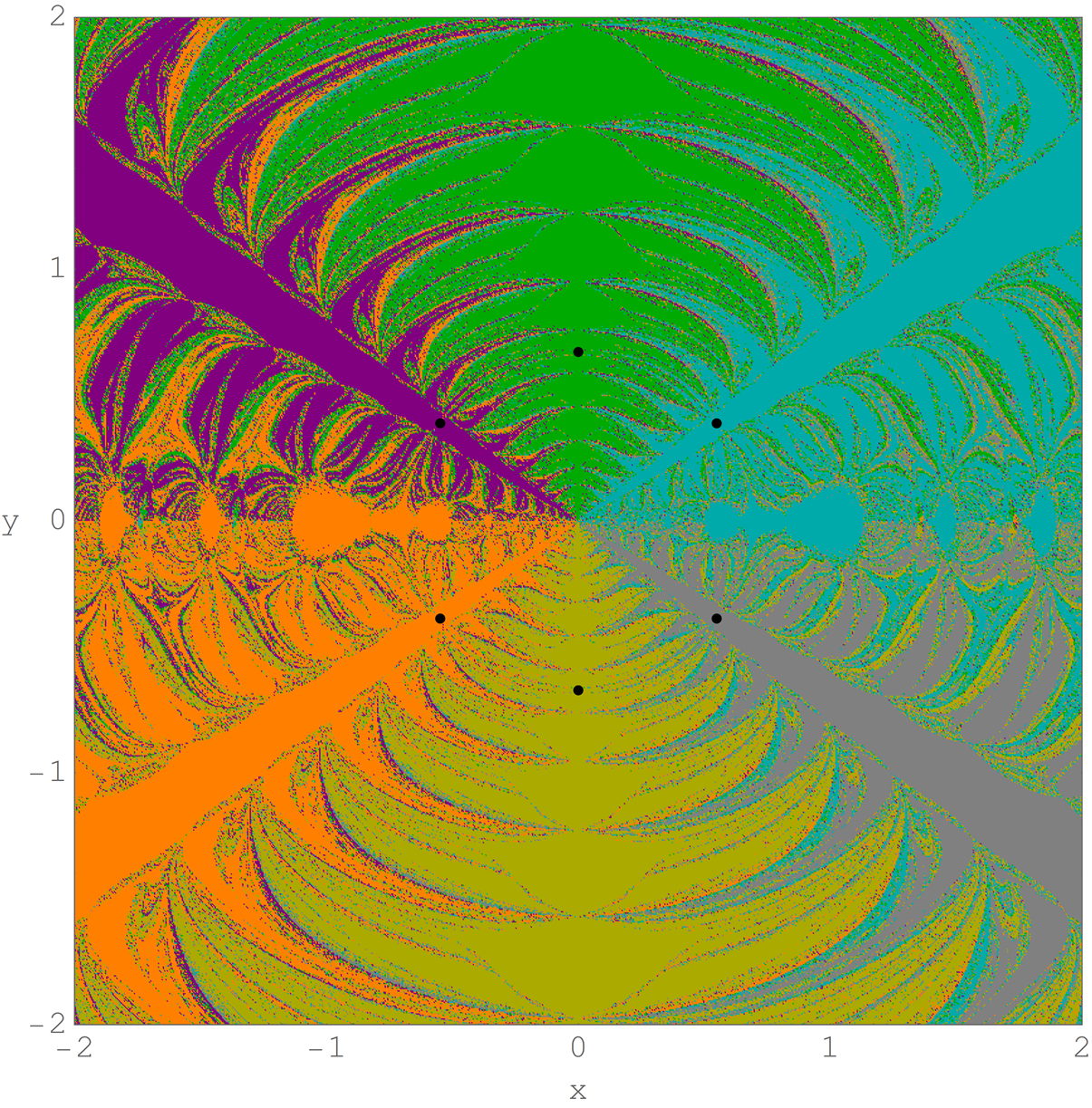}\\
\caption{The BoC linked with the libration points on $(x, y)$-plane  for  $\beta=1000$, and  (a) for $A=-0.097968$, (b) $A=-0.001$. The color codes for the BoC are as follows: $L_{xy1}^+=$\emph{teal}, $L_{xy2}^+=$\emph{purple}, $L_{xy1}^-=$\emph{gray}, $L_{xy2}^-=$\emph{orange}, $L_{y1}^+=$\emph{green} and $L_{y1}^-=$\emph{olive}.
The dots show the  positions of libration points.  (colour figure online).}
\label{Fig:Basin_3}
\end{figure*}
%%%%%%%%%%%%%%%%%%%%%%%%%%%%%%%%%%%%%%%%%%%%%%%%%%
%%%%%%%%%%%%%%%%%%%%%%%%%%%%%%%%%%%%%%%%%%%%%%%%%%
\section{The basins of convergence}
\label{BoC}
In the present manuscript a systematic study related to the basins of convergence (BoC) associated with the libration points are presented when the peripheral primaries are non-spherical in shape. The  well known method in multivariate version, i.e., Newton-Raphson method is used to solve the nonlinear equations. In addition, we will use the procedure and methodology used by \cite{zot16}  to illustrate the domain of the basins  of convergence in the in-plane case only.

The associated multivariate iterative scheme is
\begin{equation}\label{Eq:401}
 \textbf{x}_{n+1}=\textbf{x}_n-\mathbb{J}^{-1}f(\textbf{x}_n),
\end{equation}
where $f(\textbf{x}_n)$ shows the system of equations, whereas the associated inverse Jacobian matrix is given by $\mathbb{J}^{-1}$.
In our system \ref{Eq:1a}-\ref{Eq:1c}, we have three equations.

Thus, we can use the multivariate NR iterative scheme on the system:
\begin{subequations}
\begin{eqnarray}
\label{Eq:301a}
 U_x(x, y)&=&0, \\
\label{Eq:301b}
U_y(x, y)&=&0,
\end{eqnarray}
\end{subequations}
and for the configuration $(x, y)$ plane, for each coordinate, the bivariate  version of the iterative scheme are read as:
\begin{subequations}
\begin{eqnarray}
\label{Eq:304a}
x_{n+1}&=&x_n-\frac{U_{x_n}U_{y_ny_n}-U_{y_n}U_{x_ny_n}}{U_{x_nx_n}U_{y_ny_n}-U_{x_ny_n}U_{y_nx_n}},\\
\label{Eq:304b}
y_{n+1}&=&y_n+\frac{U_{x_n}U_{y_nx_n}-U_{y_n}U_{x_nx_n}}{U_{x_nx_n}U_{y_ny_n}-U_{x_ny_n}U_{y_nx_n}}.
\end{eqnarray}
\end{subequations}
In the above equations, the values of $x$ and $y$ coordinates at the $n$-th step of the iterative scheme are given by $x_n$ and $y_n$ in the Newton-Raphson method. Here, the corresponding second order partial derivatives of the potential function are represented by the subscripts of $U(x, y)$.

%%%%
The philosophy that works behind the Newton-Raphson iterative scheme is as follows: the numerical code activates when the initial condition  $(x_0, y_0)$ is provided on the plane, whereas the iterative scheme continues until an attractor (i.e., equilibrium point) is achieved, with the coveted predefined accuracy.  Whenever the specific initial condition reached to one of the attractor (i.e., libration point) of the dynamical system, we claim that for that specific initial condition, the iterative scheme converges. It can be noted that the iterative scheme does not converge equally well for each of the initial conditions, in general. The collections of all those initial conditions which converge to the same attractor compile the so-called BoC or NR basins of attraction.

It is well known that in the dynamical system such as $N$-body problem, it is not always possible to find the analytic formulae to evaluate the coordinates of positions of the equilibrium points.  In fact, for $N>3$, we do not have any analytical formulae to evaluate the position of the libration point, thus,  the one of the best means is to use the numerical methods to evaluate their positions numerically. At this point of time it is necessary to note that every numerical method strongly depends on the choice of initial conditions. Indeed, the numerical methods may converge to one of the attractors (i.e., the roots) or may need a vast number of iteration for some of the initial conditions to converge at one of the attractor while for some of the initial conditions the numerical method may trapped into an endless cycle in a periodic or aperiodic manner or may diverge to infinity. This fact leads to the conclusion that the choice of initial condition must be good so that the method may converge to one of the attractor.
The various literature concerning the iterative scheme suggest that those initial conditions need less number of the iterations which lie in the regular domain of the basins of convergence, on the other hand  the initial conditions which fall in the fractal region may need a huge number of iterations to converge at one of the attractor. The above mentioned facts give us a considerable amount of reason to examine the BoC corresponding to the equilibrium points.  Consequently, we can pick those initial conditions easily for which the iterative method need lowest number of iterations to converge at one of the specific attractor with predefined accuracy. Moreover, the most intrinsic properties of the dynamical system are unveiled by the analysis of the BoC corresponding to the libration points. Since , the equations (\ref{Eq:304a}, \ref{Eq:304b}) contain the first and second order derivatives  which  combine the dynamics of the test particle's orbit together with the corresponding stability properties and therefore, it gives a strong reason for revealing the basins of attraction in the present dynamical model.

%%%%
Recently, various scientists and researchers have deplete their time to analyze the NRBoC in different types of dynamical systems, such as the restricted problem of three bodies (e.g., \cite{dou12}, \cite{zot16}, \cite{Z17}),  the restricted problem of four bodies (e.g.,  \cite{Z16}, \cite{sur17a}, \cite{sur17b}), and the restricted problem of five bodies (e.g., \cite{ZS17}).
%%%

To illustrate the NRBoC, we deploy the following algorithm: we classify the configuration plane into dense uniform grids of $1024\times1024$ initial conditions $(x_0, y_0)$ and perform a double scan of the configuration $(x,y)$ plane.
For the present numerical computations, the maximum number $N_{max}$ of iterations allowed is set to $500$ while the predefined accuracy is set to $10^{-15}$ regarding the coordinates of the equilibrium poiints. At this point of time it is necessary to make clear that the Newton-Raphson BoC  must not be mistaken with the basins of attraction of that of dissipative systems. It should be noted that the present manuscript deals with numerical attractors and the BoC associated with them.

\subsection{In-plane basins of convergence}\label{IPBOC}
We begin our numerical analysis with the in-plane case i.e., for the case when libration points lie on $(x,y)$ plane only. To classify each nodes on the configuration plane we have used the color coded diagrams (CCDs), where each different color is associated with each pixel, as per the final stage of the associated initial conditions. In Section \ref{LP:0}, we have illustrated that there exist different type of sets of libration points for different combinations of the parameter $\beta$ and $A$. On this basis we have divided our analysis in following subsections.
\subsubsection{When libration points lie on xy-plane}\label{IPBOC-onplane}
In this subsection, we have discussed the BoC in those cases for which the libration points exist on the axes as well as on the $xy-$plane. In Fig.\ref{Fig:Basin_0}, we have illustrated the BoC for fixed value of $\beta=0.1$ and various values of parameter $A$. The configuration plane is covered by well formed BoC linked to the equilibrium  points. Moreover, the extent of BoC are infinite linked to all libration points.  We observed that the libration points linked to $L^{-}_{x1}$ and $L^{+}_{x1}$ are resemble with exotic bugs with many legs and antennas. It is obvious that the majority of area of the configuration plane are covered by BoC linked to the libration points $L^{\pm}_{x4}$ and $L^{\pm}_{y1}$. However, the basin boundaries are composed of the mixture of the initial conditions and looks like chaotic sea. As the value of parameter $A$ decreases from $-0.005$ to $-0.01$, there is very few changes in the BoC linked to libration points. However, the legs and antennas of the exotic bugs decrease significantly which shows that the domain of the BoC decrease corresponding to libration points $L^{-}_{x1}$ and $L^{+}_{x1}$.  In Fig.\ref{Fig:Basin_0}c, the BoC is depicted for large scale of the configuration plane $(x, y)$ to have the idea of the BoC on broader scale. It can be noticed that multiple wings shaped region linked to the equilibrium points $L^{\pm}_{x3}$ (magenta and brown) increases its wingspan and the boundaries of the basins looks highly chaotic, infact the area between the wings shaped region linked to $L^{\pm}_{x3}$ and $L^{\pm}_{y1}$ is highly chaotic which is composed of the mixture of various initial conditions. Therefore, it is very difficult to predict the final state of the initial conditions which falling in these chaotic reasons. As we compare the BoC with the previous panels we can notice that as the value of $A$ decreases, the libration points $L^{\pm}_{x3}$ and $L^{\pm}_{x4}$ come closer to each other respectively and consequently a major area of the domain of BoC of the configuration plane is covered by those initial conditions which converge to these libration points. The domain of the BoC corresponding to libration points $L^{\pm}_{xyi}$, $i=1,2$, looks like four lobes originating from $x-$axis. The domain of the BoC linked to these libration points increase as the value of $A$ decrease. In Fig.\ref{Fig:Basin_0}d, the BoC is illustrated for that value of $A$ for which the libration points $L^{\pm}_{x3}$ and $L^{\pm}_{x4}$ collide and annihilate completely, and consequently only ten libration points exist. We noticed that a major area of the configuration plane looks like chaotic sea, composed of the initial conditions, however the areas in the neighbourhood of the equilibrium points are regular.
If we compare  panels Fig.\ref{Fig:Basin_0}a,b, to panel Fig.\ref{Fig:Basin_0}d,  we notice that the area adjoining the exotic bugs shaped region which looks highly chaotic and appears elliptic in shape, increases significantly. We believe that the initial conditions which converge to libration points $L^{\pm}_{x3, 4}$ when these points exist, now converge randomly to any of the existing libration points which is the reason why this region turn into the chaotic sea. Infact, the regular island of BoC which converges to the equilibrium points $L^{\pm}_{x3, 4}$ turned into the chaotic sea. The zoomed view near the origin shows that a elongated  "eight" shaped region exist from $L^{-}_{x2}$ to $L^{+}_{x2}$ is also chaotic except the four regular lobes shaped basins linked to $L^{\pm}_{xyi}$, $i=1,2$.

In Fig. \ref{Fig:Basin_2}, the BoC is illustrated for fixed value of $\beta=1000$ and those values of $A$ for which there exist ten equibrium points. We noticed that in both the panels, the extent of the BoC is infinite linked to all the equilibrium points. Moreover, as the value of $A$ increases from -0.142 to -0.09804989, the BoC linked to libration points $L^{\pm}_{xyi}$, $i=1,2$, becomes more regular also the domain of the BoC linked to the equilibrium points $L^{\pm}_{xi}$, $i=1,2$, decrease. However, in both the cases a major part of the configuration plane is covered by chaotic sea composed of the various initial conditions.
\subsubsection{When libration points lie on axes only}\label{IPBOC-onaxes}
In this subsection, we will discuss the BoC in those case for which the libration points lie only on the axes.  In Fig. \ref{Fig:Basin_1}, we have discussed the BoC for two different values of $\beta=0.1, 1000$, and different values of $A$.  Fig. \ref{Fig:Basin_1}a, when $\beta=1000$ and $A=0.025$, there exist four equilibrium points on $x-$axis while two equilibrium points on $y-$axis. We noticed that the configuration plane is covered by well-formed BoC which extended to infinity. However, the majority of the area is occupied by domain of the BoC linked to the equilibrium points $L_{y1}^{\pm}$, which looks like the multiple butterfly wings. In addition, the domain of BoC which exists in the vicinity of the equilibrium points  $L_{x1}^{\pm}$  looks like circular island and the domain of BoC linked with the collinear equilibrium points $L_{x2}^{\pm}$ looks like lobes originate from origin along the $x-$ axis. Further, as the  value of the oblateness parameter $A=0.75$ (see Fig. \ref{Fig:Basin_1}b) increases, it is seen that the butterfly wings shaped regions shrink in the vicinity of $x-$axis and consequently domain of the BoC associated with the  equilibrium points $L_{x2}^{\pm}$ increases whereas there is negligible change in the circular island shaped  BoC linked with $L_{x1}^{\pm}$.

Fig. \ref{Fig:Basin_1}c, the BoC is  illustrated for $\beta=0.1$ and $A=-0.215$. It is seen that the basin of convergence associated with the equilibrium points $L_{x1}^{\pm}$ (see purple \& green colors)  looks like exotic bugs with many legs and antenna however, the BoC  linked to libration points $L_{x2}^{\pm}$ the two exotic bugs shaped region corresponding to  each libration points (shown in cyan and olive color) exist in which one exists in the vicinity of the libration points and other exists adjacent to the domain of BoC linked to the equilibrium points $L_{x1}^{\pm}$ and wings and antennas of these exotic bugs are very noisy.  It is observed that these wings and antenna are chaotic mixture of the initial conditions which converge to different attractors and consequently it is almost impossible to predict that to which attractors these initial conditions are converging.
\subsubsection{When collinear libration points do not exist}\label{IPBOC-onaxes}
In this subsection, we have illustrated the BoC for fixed value of $\beta$ and various values of $A$ for which there exist only non-collinear libration points (see Fig. \ref{Fig:Basin_3}). The extent of the domain of BoC linked to each of the libration point is also infinite in these cases. When the value of $A= âˆ’0.097968$, majority of the areas of the configuration plane is covered by domain of BoC linked to the libration points  $L_{y1}^{\pm}$. The well formed domain of the BoC linked to the libration points  $L_{xyi}^{\pm}$, $i=1,2,$ exist in the vicinity of the $x-$axis. However, the entire configuration plane look like a chaotic sea of initial conditions except for those BoC which are associate to $L_{xyi}^{\pm}$ (see Fig. \ref{Fig:Basin_3}a). The BoC changes drastically as the value of the $A$ increase to $A=-0.001$.  The regular domain of the BoC linked to the libration points $L_{xyi}^{\pm}$ increases significantly and consequently the domain of the BoC linked to $L_{y1}^{\pm}$ decreases. However, the basins boundaries are very noisy and, indeed, it is impossible to predict the final state of the initial condition falling inside these basins boundaries even after a sufficient number of iterations.
%%%%%%%%%%%%%%%%%%%%%%%%%%%%%%%%%%%%%%%%%%%%
\section{Concluding Remarks}
\label{CR:0}
In the present problem, the collinear restricted four body problem has been investigated where the peripheral primaries are non-spherical in shape. In particularly, the shape of the peripheral primaries are either oblate or prolate spheroid.

It should be emphasized that a numerical investigation is performed in such a thorough and systematic manner to unveil the effect of the parameters $\beta$ and  $A$ on the position and stability of libration points, regions of possible motion and on the topology of BoC by applying the NR iterative scheme that all the illustrated results are novel, whereas they enhance  substantially to our  knowledge related to the evolution of the libration points.

The most important outcomes of our numerical analysis  are listed below:
\begin{itemize}
  \item [$\bullet$] The existence as well as the number of collinear libration points strongly depend on the particular values of parameters $\beta$ and $A$. For $\beta=0.1$, there exist eight collinear libration points for $A\in(-0.039515, 0)$ while it reduces to four when $A\in (-0.259259, -0.039515)$ and (0, 1), in addition, for $\beta=1000$, there exist four collinear equilibrium points when $A\in (-0.166687, $ $-0.09805989)$ and  $[0,1)$ whereas there exist no collinear libration point for $A\in (-0.09805989, 0)$. It is further noticed that there exist two type of non-collinear libration points, i.e., the libration points which lie on $(x, y)$ plane and the libration points which lie on $y-$axis for the considered values of $\beta=0.1, 1000$ and varying values of the  parameter $A$.
  \item [$\bullet$] The movement of the libration points which lie on $y-$ axis remains toward the central primary as the value of the parameter $A$ increases. The libration points $L_{xy1}^{\pm}$  and  $L_{xy2}^{\pm}$  originate in the vicinity of the libration point $L_{x2}^{\pm}$ and move far from the $x-$axis along the arc shaped trajectories and again turned to annihilate in the neighbourhood of the peripheral primaries for $A=0$.
  \item[$\bullet$] The stability analysis for the in-plane libration points suggests that only some of the libration points which lie on  the $x-$axis are linearly stable for a particular value of $\beta$ and particular range of the parameter $A$. Whereas none of the non-collinear libration points are found linearly stable for any permissible value of the parameter $A$ for the studied value of $\beta$.
  \item[$\bullet$] The parametric evolution of the regions of possible motion in the in-plane case is illustrated and found that as the value of the Jacobian constant increases, the regions of the possible motion, where the test particle is  free to move, decrease. Moreover, the region of forbidden motion decreases in the in-plane case when the value of $A$ decreases.
   \item[$\bullet$]  The evolution of the BoC  as the function of the parameter $A$ is presented to determine the effect of the parameter $A$ on their topology. It is observed that in all the cases the extent of the domain of BoC linked to the libration points is infinite. The basins boundaries are always chaotic in  nature which are mainly composed of those type of initial conditions which converge to any of the equilibrium points randomly.
\end{itemize}
We have used the latest version $12.0$ of the Mathematica$^\circledR$ to perform all numerical as well as graphical illustrations. We hope that the presented numerical analysis and the obtained results to be useful in the real world  where the primaries are celestial bodies which are in collinear  configuration. In future work it is interesting to unveil that how the topology of the BoC linked to the out-of-plane equilibrium points, if exists, is changed with the change when the value of parameters $\beta$ and $A$ changes.
\section*{Compliance with Ethical Standards}
\begin{itemize}
\item[-]Funding: The author, Rajiv Aggarwal,  has received the research grant by Department of Science and Technology, New Delhi, India.
\item[-]Conflict of interest: The authors declare that they have no conflict of interest.
\end{itemize}
\section*{Acknowledgements}
This work is funded by \textbf{Department of Science and Technology, India},  under project scheme \textbf{MATRICS (MTR/2018/000442)}.

%%%%%%%%%%%%%%%%%%%%%%%%%%%%%%%%%%%%%%%%%%%%%%%%%%

%%%%%%%%%%%%%%%%% APPENDICES %%%%%%%%%%%%%%%%%%%%%

% Don't change these lines
%\bsp	% typesetting comment
\label{lastpage}
\end{document}